\documentclass[]{interact}
\usepackage{amssymb}
\usepackage{amsmath}
\usepackage{multirow}
\usepackage[table]{xcolor}
\usepackage{graphicx}
\usepackage{graphics}
\usepackage{subfig}
\usepackage{float}
\usepackage{caption}
\usepackage{todonotes}
\usepackage{url}

\theoremstyle{plain}

\theoremstyle{definition}

\theoremstyle{remark}

\def\beq{\begin{equation}}
\def\eeq{\end{equation}}
\def\bqn{\begin{eqnarray}}
\def\eqn{\end{eqnarray}}

\def \ee {\begin{equation}}
\def \eee {\end{equation}}
\def \eqe {\begin{eqnarray}}
\def \eqee {\end{eqnarray}}

\definecolor{forestgreen}{rgb}{0.0,0.5,0.1}
\definecolor{plum}{RGB}{142,69,133}
\definecolor{darkcyan}{rgb}{0, 0.4, 0.45}

\usepackage[left]{lineno}

\title{Social decision-making driven by artistic explore-exploit tension}
\author{\name{Kayhan \"{O}zcimder\textsuperscript{a}, Biswadip Dey\textsuperscript{a}, Alessio Franci\textsuperscript{b}, Rebecca Lazier\textsuperscript{c}, Daniel Trueman\textsuperscript{d}, and Naomi Ehrich Leonard\textsuperscript{a}
\thanks{Corresponding author: N.~E. Leonard. Email: naomi@princeton.edu, Telephone: +1 609/258-5129} 
\thanks{Please cite as follows: K. \"{O}zcimder, B. Dey, A. Franci, R. Lazier, D. Trueman \& N. E. Leonard (2018): Social decision--making driven by artistic explore--exploit
tension, {\em Interdisciplinary Science Reviews}, DOI: 10.1080/03080188.2018.1544806} 
}
\affil{\textsuperscript{a}Department of Mechanical and Aerospace Engineering, Princeton University, Princeton, NJ, USA; \\ \textsuperscript{b}Department of Mathematics, National Autonomous University of Mexico, Mexico City, Mexico; \\ \textsuperscript{c}Program in Dance, Lewis Center for the Arts, Princeton University, Princeton, NJ, USA; \\ \textsuperscript{d}Department of Music, Princeton University, Princeton, NJ, USA}
}

\date{}

\begin{document}
\maketitle

\begin{abstract}
We studied social decision-making in the rule-based improvisational dance {\em There Might Be Others}, where dancers make in-the-moment compositional choices.  Rehearsals provided a natural test-bed with communication restricted to non-verbal cues.  We observed a key artistic explore-exploit tension in which the dancers switched between exploitation of existing artistic opportunities and riskier exploration of new ones. We investigated how the rules influenced the dynamics using rehearsals together with a model generalized from evolutionary dynamics. We tuned the rules to heighten the tension and modeled nonlinear fitness and feedback dynamics for mutation rate to capture the observed temporal phasing of the dancers' exploration-versus-exploitation.  Using bifurcation analysis, we identified key controls of the tension and showed how they could shape the decision-making dynamics of the model much like turning a ``dial'' in the instructions to the dancers could shape the dance. The investigation became an integral part of the development of the dance.

\end{abstract}


\section{Introduction}
Social decision-making enables group efforts that are neither fully scripted nor centrally controlled \cite{Dyer2009}, so predicting group behavior requires understanding what drives individual choices \cite{Sanfey2007}.  Social decision-making is studied in various research communities, including social choice theory \cite{Arrow1951,Sen1970,Saari2001}, which combines social ethics with voting theory, social neuroeconomics \cite{Sanfey2007,FehrCamerer2007,LeeNatureNeuro2008}, which joins game theory  with psychology and neuroscience, and collective animal behavior \cite{SeeleyBuhrman1999,ConradtRoper2003,Couzin2005,MeunierMonkeys2006,Leonard2012}. In network science, researchers examine the role of network structure in decentralized decision-making groups \cite{OlfatiFaxMurray2007,Fefferman2015,LazerFriedman2007,MasonWatts2009,LandgrenCDC2016}.

A fundamental consideration in settings where options yield rewards is how decision-makers make choices that balance exploitation of options with well-known rewards with the riskier, but possibly advantageous, exploration of options with poorly known rewards. In reward-based decision-making, the decision-maker chooses from a set of options and receives a reward associated with the chosen option. Rewards may be uncertain and variable, and so the decision-maker seeking to maximize reward faces a dilemma between choosing an option that's known to yield reasonably high reward (exploitation) versus choosing an option for which they have little information but that could yield even higher reward (exploration). Exploitation and exploration are in tension because a decision-maker who only exploits will not get the new information that comes from exploring, and a decision-maker who only explores will not leverage the new information received. 

When the decision-maker is allowed a sequence of reward-bearing choices over time, the explore-exploit tension changes with each choice. For example, consider choosing a restaurant for dinner in a city where you are visiting for an extended time. Suppose you have an outstanding meal at restaurant A on the first night. Exploiting the newly gained information, you return the second night but have merely a decent meal. Then, on the third night, you wonder if you should exploit your two data points and go back to restaurant A for what you can anticipate will be something in between decent and outstanding. Or do you explore a restaurant B hoping to find a consistently outstanding meal? 

A large literature addresses the explore-exploit tension for single decision-makers and provides algorithms for choices that reliably optimize the accumulation of reward over time \cite{TLL-HR:85,PA-NCB-PF:02,EK-OC-AG:12}. The tension is examined in a wide range of contexts, including control of attention in the brain \cite{CohenExplo2007}, allocation of treatments in clinical trials \cite{Villar15}, and animal foraging in a patchy environment \cite{Krebs1978}. 
The tension has more recently been studied for groups of interacting decision-makers, using simulation \cite{LazerFriedman2007}, experiment \cite{MasonWatts2009}, and model-based analysis \cite{LandgrenCDC2016,kalathil2014decentralized,KollaJG16}. In these group decision-making settings, individuals make their own choices among options, but they are influenced by social interactions that involve observations and communications of the choices or evaluations that others in the group make.

The studies of decision-making groups raise important open questions about how factors in the social interactions influence the explore-exploit tension and the group's decision-making dynamics. How important is the clarity of the communication of choices and rewards? Does it matter if interactions are with experts or with novices?  What is the role of the network structure of the interactions, i.e., who can observe or communicate with whom?  \cite{MasonWatts2009} ran an experiment in which each participant was asked to make a sequence of choices of location on a computer generated map in order to find the deepest oil well. After every choice, each participant could see their  reward (how deep the oil at the location selected) as well as the choices and rewards of some of the other participants. The data showed that networks (as compared to solitary decision-making) suppressed individual exploration, i.e., individuals copied one another more often than trying out a new location.  However, when someone in the group made a highly rewarding choice, the whole group benefited.  

The authors cautioned that their results were likely dependent on the structure of the environment, where there was only one location with a deep oil well and thus little information to be gained by exploring the rest of the map.  The average reward at each location also remained  constant, which has been the case in most settings examined in the literature.  When average rewards don't change, decision-makers can ultimately stop exploring once they have enough information to find the best option. These observations illustrate the importance of understanding the influence of the environment on the explore-exploit tension and the group's decision-making dynamics. For instance, if the rewards change, how frequently will the decision-maker explore? What happens if environmental conditions create urgency in decision-making? How does decision-making change if sticking with the current option is easy and  choosing a new one is costly? What if the reward associated with an option decreases if a decision-maker sticks with it too long?  Or what if the reward is diminished when too many individuals select the same option at the same time? 

In the present paper, we report on a novel, generative investigation of {\em There Might Be Others} (TMBO), an open choreographic work where performers collectively create the piece in real-time negotiating a catalog of defined movement ``modules'' with a set of performer instructions and governing rules. TMBO builds on the tradition of open scores and improvisational works wherein the performers compose the work in performance within a set of rules and contingencies. The artistic quality of TMBO unfolds as the dancers experiment with relationships, timing, space, and groupings and collectively work to ensure unpredictability. Through their decisions balancing how, when, and where to perform the defined vocabulary, the dancers create beautiful moments of juxtaposition, complex groupings, and dynamic shifts in tempo. 

Our investigation arose out of an art and science collaboration aimed at finding principled ways to influence the creative process in social, rule-based art-making. As it turned out, the investigation led to this and much more, including a fresh perspective on the development of the piece and the means to examine open scientific questions.  TMBO provided a rich opportunity for studying social decision-making in a creative endeavor by highly trained artists, with rehearsals serving as a natural test-bed and dynamics well-suited to mathematical modeling and analysis. In TMBO, the dancers' decisions carry both artistic rewards and risks. Recognizing this, we identified an explore-exploit tension driving the artistic choices of the dancers, and a connection between the tension and the rules of the dance. We defined choosing to ``exploit'' as joining a module currently being danced, and choosing to ``explore'' as introducing a module that has not yet been danced. These are in tension, and yet the resulting dynamics are quite different from the standard explore-exploit dynamics in which average rewards are stationary. In TMBO, rewards can change depending on the sequence of choices, the changing environment, and the dancers' artistic sensibilities. For example, a module that initially was highly rewarding might lose its appeal if it is danced for too long; one or more dancers might then seek to explore something new. The result is a richly varied dynamic involving periodic switching between exploration and exploitation.

To investigate the mechanisms at play and the opportunities for design, we made a systematic examination of the rules, environmental context, explore-exploit tension and the overall effect on the decision-making dynamics and the dance. Our approach integrated ongoing rehearsals with the TMBO dancers and analysis of a low-dimensional mathematical model of the explore-exploit tension, which we generalized from evolutionary dynamics \cite{Burger_98}. In rehearsal we modified rules and recorded the resulting behavior of the dancers. With the model we studied the sensitivity of the explore-exploit dynamics with respect to parameters in the modeled rules and context. Our observations in rehearsal led to refinements in the model.  And the model provided an off-line tool for examining mechanisms and for investigating new opportunities before trying them in rehearsal and designing them into practice. 
We learned much about the driving forces behind the underlying switching dynamics of the dance and how these dynamics could be shaped with subtle modification to instructions. Along the way, the language of our explore-exploit study became our baseline language for making TMBO. 

In this paper, we describe our investigation of the artistic explore-exploit tension in TMBO, its driving influence on the group's decision-making dynamics, and the unfolding of the dance. We reveal how the investigation itself became an integral part of the development of the dance. And we show how the results of the investigation, including the dance and the model, provide new ways of understanding and shaping these and related social decision-making dynamics.

%
%
\section{Structure and rules of the dance}

TMBO was conceived by choreographer Rebecca Lazier who was inspired by the musical piece {\em In C} composed by Terry Riley in 1964. Mid-development, Rebecca invited composer Dan Trueman and engineering professor Naomi Leonard to join the project; they were examining emergence in structured improvisation and were inspired by both {\em In C} and the art-science project {\em Flock Logic} \cite{flocklogic}. In {\em In C} musicians make compositional choices as they play an ordered sequence of fifty-three musical melodic patterns \cite{In_C_orig, In_C_Ref}. All of these patterns fit on a single page, and the instructions for the piece are similarly concise. {\em In C} is both participatory and emergent, relying on collective musical decision-making from a group of expert performers to create a performance that is always recognizable but also unique; it is widely considered one of the most influential pieces of 20th-century concert music, and the beginning of musical Minimalism \cite{Carl2009}.

Similarly, TMBO is defined by a catalog of forty-four movement modules (in analogy to the musical melodic patterns) and performing instructions that lay out the choreographer's artistic objectives and rules for how a group of dancers can manipulate the modules in performance \cite{lazier2016there}.  The objectives and rules also apply to how a group of dancers can order the modules in performance, since, unlike in the Riley piece, no prescribed order is set in the TMBO score. Riley instructs the players to repeat each melodic pattern as many times as they wish before moving on to the next, to proceed in order, and to listen to the group to stay within three patterns of one another.  In TMBO, performers use these general rules, except that dancers are responsible for the compositional tone of the work by selecting which module should come next based on the needs of the piece in the moment. Once a module is introduced in the dance, the order is then set forth and followed by the entire cast. However, no one can lead two modules in a row, as the leadership must be shared by the group.  Using rules and games to create dance has a long and rich history \cite{Clemente1990}. There is also a growing body of research on cognition and the distributed choreographic process \cite{Kirsh2011}. In \cite{Amaral2005}, the mechanisms of self-assembly of teams in creative endeavors are examined.
\begin{figure}[t!]
\begin{center}
\includegraphics[width=0.75\textwidth]{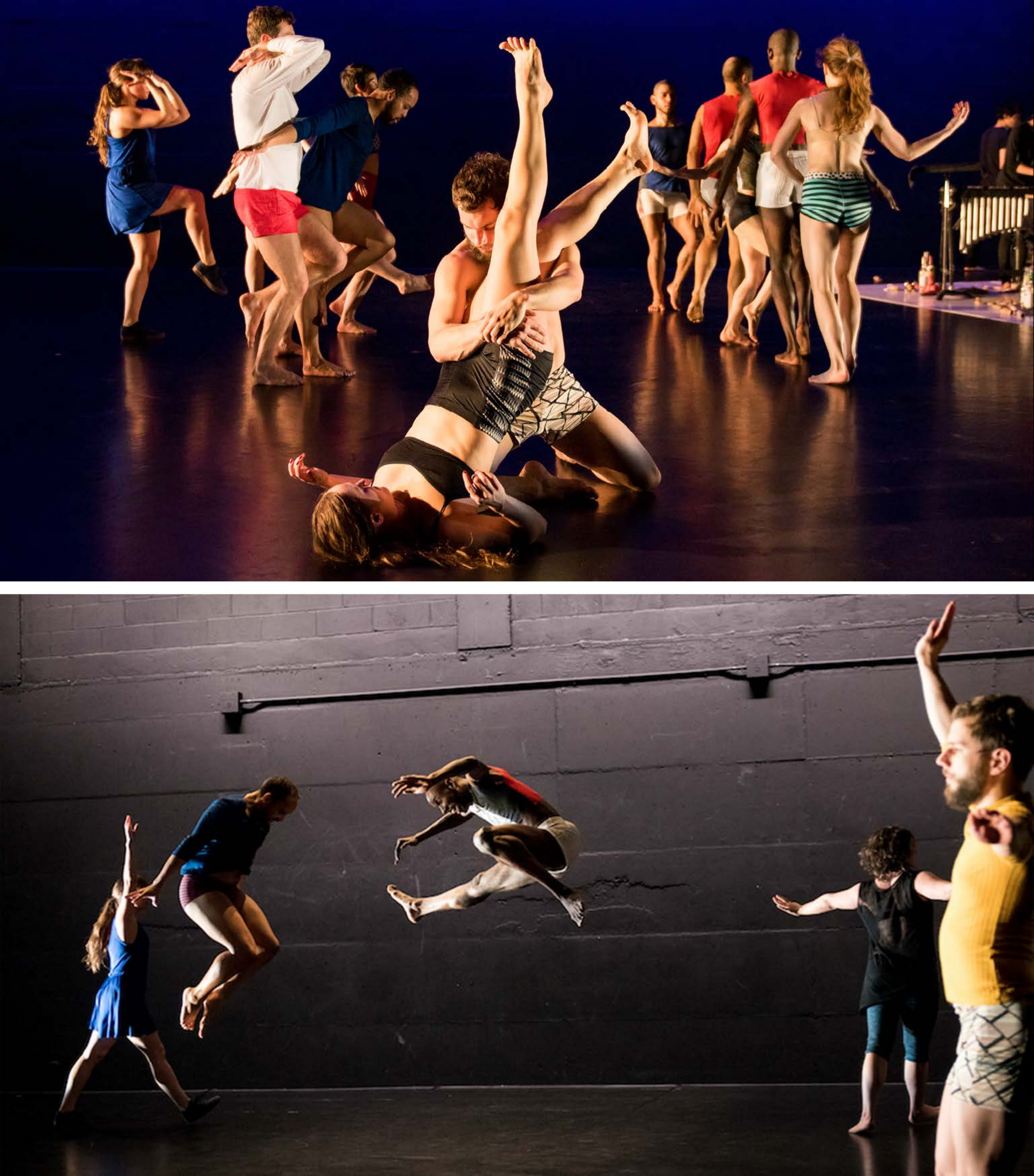}
\end{center}
\caption{Photos from a TMBO performance at New York Live Arts, New York City, in March 2016. In the top photo, several dancers on the left side perform the module {\em Flamingo} and several dancers on the right side perform the module {\em Folk}, while a pair in the foreground perform the module {\em Box and Drag}. In the bottom photo, three dancers on the sides perform the module {\em Arms} and two dancers in the center perform the module {\em Jump Bean}. Photo credit: Ian Douglas.}
\label{Figure_1}
\end{figure}

TMBO's forty-four modules include a balance of composed movement, gestural ideas, and tasks. Some modules are boisterous, such as {\em Jump Bean}, where the instruction is to bounce and interact with fellow dancers, others meditative, such as {\em Oar}, where the body is gently shifting forward and backward while one hand swoops in a paddle-like action. There are complex phrases where performers will join in unison, jumping patterns that slice through the space, and single actions like {\em Human Pile} where performers pile on top of each other in stillness.  The aesthetic influences range from ballet, contemporary, and modern dance to hip hop and various folk forms.  The modules are the vocabulary of TMBO and while each is defined and recognizable, they can be adapted according to rules for each module in order to fulfill the compositional needs of the piece in the moment.  In devising the modules, we learned that they need to have dynamic, rhythmic, expressive, spatial, and stylistic range to support an aesthetic of the work that is multi-faceted and unpredictable.  Each dancer must memorize the catalog of modules and be able to not only select the one most suited for the moment but also be willing to continually develop the performance of the modules in new ways with each iteration. Fig.~\ref{Figure_1} shows two photos from a performance of TMBO at New York Live Arts, New York City, in March 2016.

TMBO starts with dancers gradually joining the performance of a single module and proceeds with dancers introducing modules.  As a new module is introduced the order is then set and each performer must proceed in the order modules appeared; in the general rules they cannot skip or switch modules.  Any dancer can introduce a new module at any time, although no more than one module can be introduced at the same time nor can a dancer lead two modules in a row; it is imperative that leadership shift among the entire population. Once a module has been introduced, danced, and abandoned, it cannot be re-introduced. Modules can be juxtaposed by different subgroups of dancers performing different modules simultaneously. However, the dancers are restricted to a limited number of modules at a time; in the general rules the limit is three modules at a time.  Dancers can experiment with and innovate variations of any current module at any time, as long as they respect the rules and keep modules recognizable to the cast to ensure clear communication. The dance is over after a fixed time or when all the modules have been performed, whichever comes first.  Because none of the dancers' decisions are made in advance and they are instructed to never repeat what they have done before, each rehearsal and performance brings fresh choices of sequencing, timing, counterpoint, relationships, and variation of modules.

By design, the dancers retain considerable creative freedom: they choose which and when, as well as where and how, modules are performed to meet artistic goals for juxtaposition, unpredictability, and dynamic pacing. Social interactions are a priority: the dancers should clearly communicate their choices through their movement and other non-verbal cues, keep up frequent observation of the movement choices of the other dancers, and make their own choices in response to what the other dancers are doing. For example, one artistic objective seeks texture and richness through juxtaposition of modules, and this can only be achieved through coordinated choices among the dancers, such as when a fraction of the group chooses to perform {\em Arms} while the rest performs {\em Jump Bean} as in the bottom photo of Fig.~\ref{Figure_1}. Another artistic objective seeks recurring moments of surprise, each defined by the introduction of a new module. This too can only be achieved through coordinated choices among dancers, since only one dancer can introduce a module at a time, and the limit on the number of current modules cannot be exceeded.  

Through direct interactions with and observations of dancers rehearsing and performing TMBO, we discovered that the social decision-making during TMBO is driven in large part by an explore-exploit tension in which rewards are artistic in nature.  By choosing to experiment with an existing module, the dancer exploits a known option.  And by choosing to introduce a new module, the dancer takes the riskier step and explores a less predictable option.   Both are creative choices and can add artistically to the performance.  We found that a key to how the dancers trade off exploiting and exploring is the performance rule that limits the number of modules that should be danced at a time. Originally, a limit of three modules was loosely applied, meaning that the dancers would sometimes add a fourth module. However, if the limit is made strict and there are already three modules being danced, a dancer cannot add a new module until all the dancers coalesce into two modules.  In this way exploiting and exploring are strongly in tension for the group: {\em either} the dancers experiment with all the current modules {\em or} they complete a current module so that one dancer can introduce a new module.  The tension exists in principle for any limit on number of current modules; however, we found that the lower the limit the greater the tension. So for our investigation, we set the strict limit to be two modules at a time, which means a dancer can add a new module only when all dancers have converged on a single module.  

The musical component of TMBO operates similarly to the dance. The musicians have their own set of modules, and while they are conceptually paired with the dance modules, in performance the musicians set their own module order and may not be lined up with the dancers. A full exploration of the musical component of TMBO is beyond the scope of this paper, and our research focus here was on the dance; in the future, it would be of interest to pursue a similar exploration of the musical decision-making dynamics, and how the musical and dance dynamics might interact. 

%
%
%

%
%
\section{Studio rehearsals}

We made a systematic investigation of the influence of changes in performance rules and context on the dancers' decision-making dynamics and the dance, during a rehearsal 
in the Patricia and Ward Hagan '48 Studio, Lewis Center for the Arts, Princeton University 
on July 25, 2015. Because the aim was to find principled ways to modify the  structure to influence the choreographic development of the dance, the Institutional Research Board (IRB) 
at Princeton University 
did not require that we seek IRB approval for the study reported in this paper.

Our systematic study focused on three abbreviated run-throughs of TMBO, each lasting nine to ten minutes, rather than the full sixty minutes of a TMBO performance. The catalog of modules available to the dancers was also reduced from forty-four to nine modules. The dancers were instructed to start with the module {\em Clapping}, which was one of the nine in the catalog for each run-through. Nine professional dancers participated in the rehearsal: Rhonda Baker, Simon Courchel, Natalie Green, Raja Feather Kelly, Cori Kresge, Christopher Ralph, Tan Temel, Sa\'ul Ulerio, and Shayla Vie-Jenkins. All nine dancers took part in Run-through 1, but only eight of the dancers took part in Run-throughs 2 and 3 since Sa\'ul Ulerio had to leave early. 
For all three run-throughs we fixed the following performance rules, slightly modified from the general rules:
\begin{enumerate}
\item any dancer can introduce a new module at any time
\item a module cannot be chosen if it has come and gone
\item any dancer can switch to any current module at any time
\item any dancer can skip a module 
\item no more than two modules can be danced at a time. 
\end{enumerate}
Rules (1) and (2) are unchanged from the general rules: only one module can be introduced at a time and the same dancer should not introduce two modules in a row.  Rules (3) and (4) are new relative to the general rules where the set order is to be followed.  Rule (5) is a modification of the original loose enforcement of a maximum of three modules at a time to a strict enforcement of a maximum of two modules at a time.

Before they performed the run-throughs, the dancers were not informed of the motivation for the modified rules, the model, nor the investigation of an explore-exploit tension.  Each run-through was recorded with two high definition video cameras angled to capture the whole stage.

\subsection{Run-through 1} 
\begin{figure}[t!]
\centering
\includegraphics[width=0.95\textwidth]{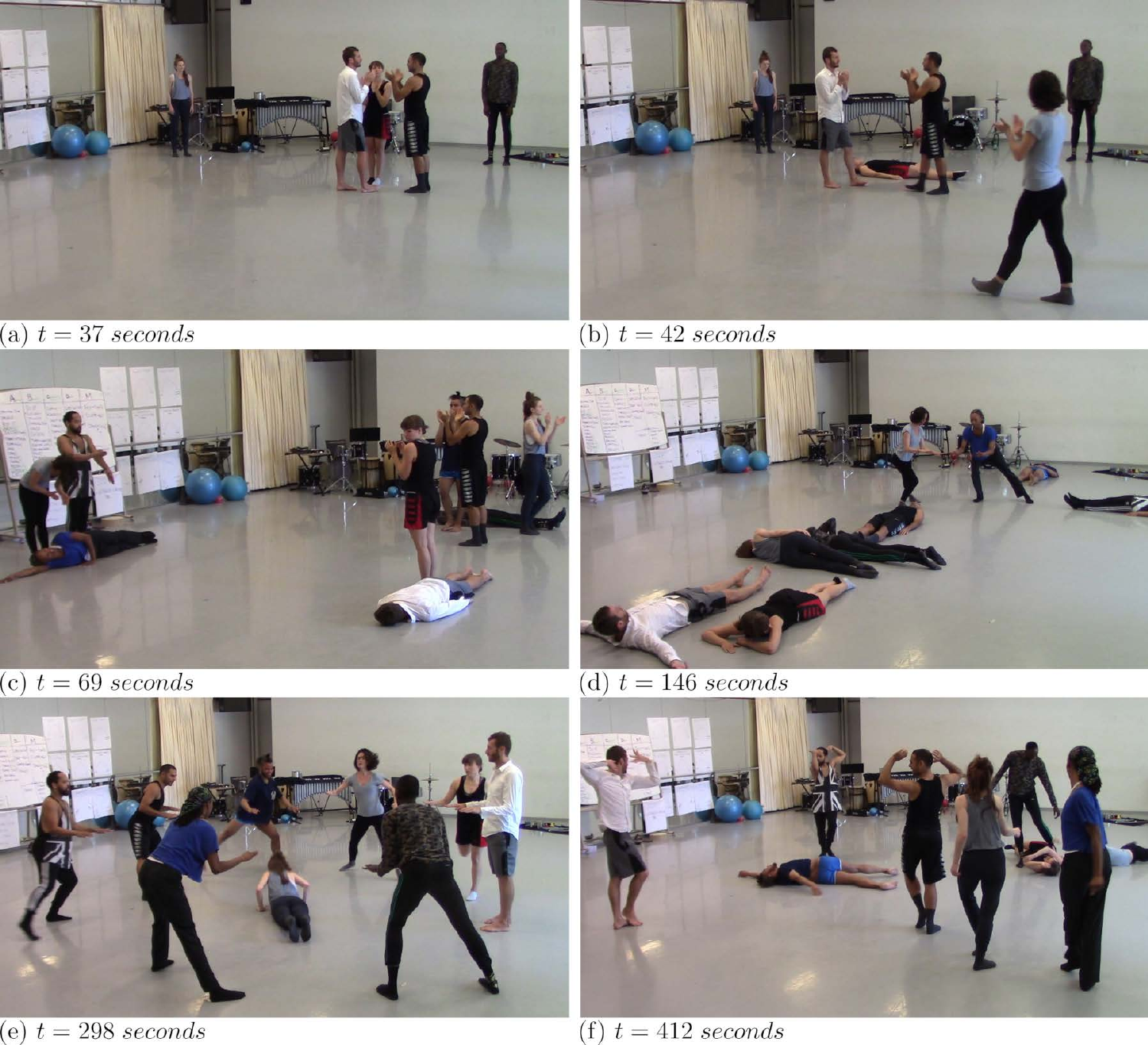}
\caption{Run-through 1 
}
\label{Figure_2}
\end{figure}

In Run-through 1, the catalog of available modules included the following nine: {\em Clapping, Flamingo, Chain, Whip It, Crawl and Sing, Trigger, Do Op, Lay Down and Get Up, Drop and Roll}.  The run-through began with dancers starting the module {\em Clapping}.  A snapshot at 37 seconds into Run-through 1 is shown in Fig.~\ref{Figure_2}(a).  In the snapshot, three dancers, Simon, Natalie, and Tan, performed {\em Clapping} facing one another in the center of the space; Rhonda and Raja can be seen on the edges of the rehearsal space not participating in {\em Clapping}, although Raja had briefly been clapping.

In the snapshot of Fig.~\ref{Figure_2}(b), at 42 seconds into Run-through 1, Cori can be seen walking and clapping. Natalie, who is lying down on her back, has clearly introduced the module {\em Lay Down and Get Up}. Since there was only one active module, {\em Clapping}, any dancer could have decided to introduce a new module among the remaining eight in the catalog at any moment, but Natalie chose {\em Lay Down and Get Up}. At this point in time, since there were two active modules, {\em Clapping} and {\em Lay Down and Get Up}, no new module could have been introduced.

In the snapshot of Fig.~\ref{Figure_2}(c), at 69 seconds into Run-through 1, the dancers were distributed over the two active modules, {\em Clapping} and {\em Lay Down and Get Up}, although  Natalie had switched back to {\em Clapping}. Cori, Chris, Rhonda, Sa\'ul, and Tan were also performing {\em Clapping}. Note that Chris (second from the left) had begun to modify {\em Clapping}; here he can be seen clapping with his arms rather than his hands. Natalie was also modifying {\em Clapping} by moving her shoulders to the beat. Shayla, lying on her side, Simon, lying on his stomach, and Raja, lying on his back, were all performing {\em Lay Down and Get Up}.  

At 146 seconds into Run-through 1, when all the dancers coalesced, at least momentarily into {\em Lay Down and Get Up}, Cori introduced the module {\em Whip It} and Shayla almost immediately joined her. The snapshot of Fig.~\ref{Figure_2}(d) shows the dancers at just this moment, with Cori and Shayla performing {\em Whip It} and the rest of the dancers evolving {\em Lay Down and Get Up}.  Since {\em Clapping} had come and gone, the dancers were no longer allowed to perform it for the remainder of Run-through 1.

By 275 seconds into Run-through 1, eight of the dancers were performing {\em Whip  It} and only one dancer (Rhonda) remained with {\em Lay Down and Get Up}. A snapshot at 298 seconds into Run-through 1 is shown in Fig.~\ref{Figure_2}(e). This lasted until around 335 seconds when Rhonda finally joined {\em Whip It}. The module {\em Do Op} was then immediately introduced. {\em Whip It} was over shortly thereafter and {\em Drop and Roll} was introduced. A snapshot at 412 seconds into Run-through 1 is shown in Fig.~\ref{Figure_2}(f), where six dancers can be seen performing {\em Do Op} and three dancers can be seen (mid-roll on the floor) performing {\em Drop and Roll}. The dancers remained with {\em Do Op} and {\em Drop and Roll} until Run-through 1 ended at 540 seconds (nine minutes). By the end, all nine dancers were performing {\em Do Op}.  Although they were available to the dancers, the four modules {\em Flamingo, Chain, Crawl and Sing}, and {\em Trigger} were not selected during Run-through 1.

\subsection{Run-through 2}

In Run-through 2, we added to the instructions of Run-through 1 by asking the dancers to be more impulsive in their choices, i.e., to {\em increase} their tendency to switch modules. The intent was to examine switching tendency as a possible design ``dial'' that could be cued to the whole group of dancers during a run-through, for example, by signaling with the beat of a drum. Switching tendency was identified as a candidate design dial since it was revealed to be associated to critical parameters in the analysis of the mathematical model as described below. For Run-through 2 the catalog of available modules included the following nine: {\em Clapping, Arms, Box, Bravo, Bumper Body, Chris, Folk, Kiss, Romper Room}. 
\begin{figure}[t!]
\centering
\includegraphics[width=0.95\textwidth]{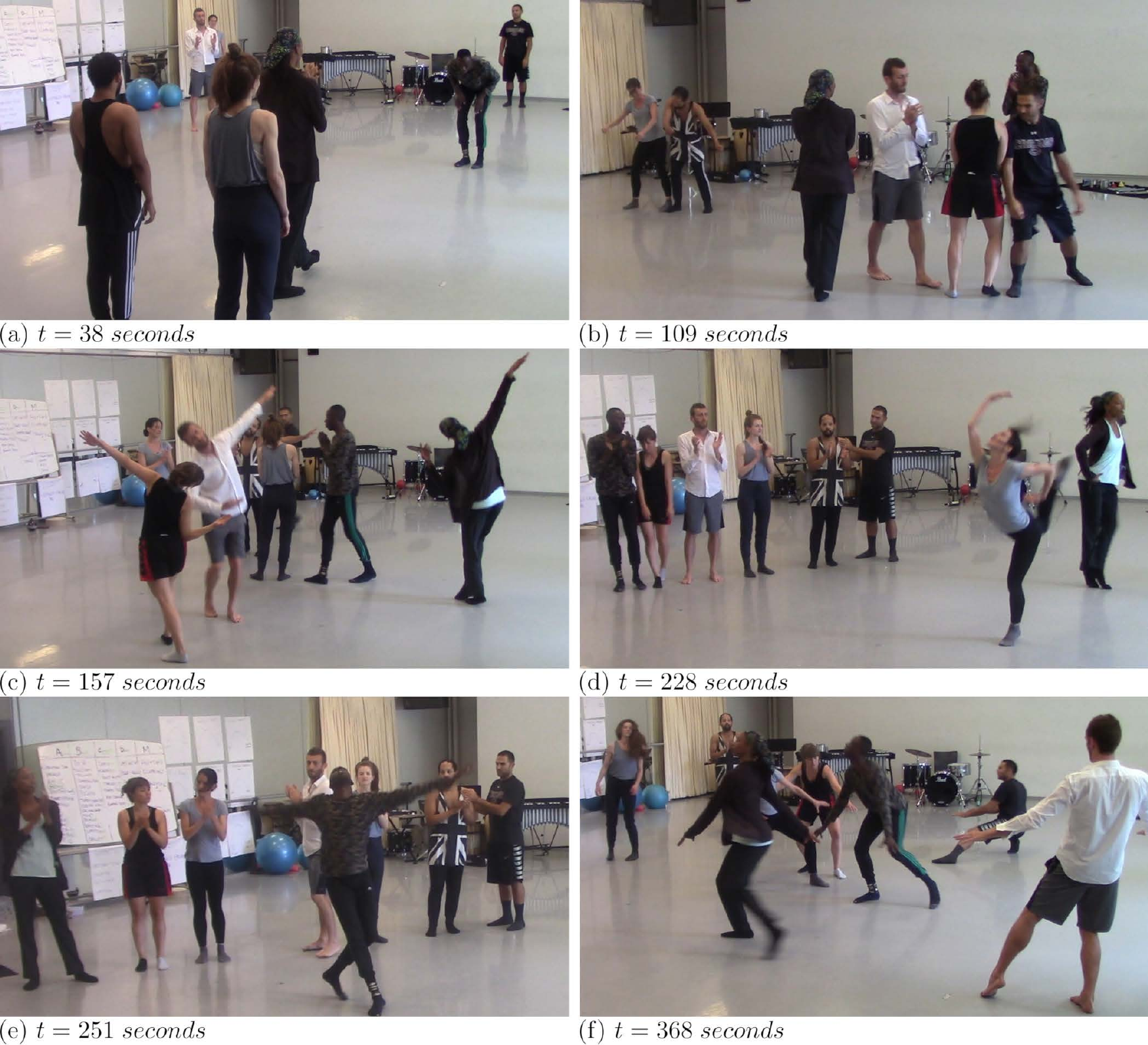}
\caption{Run-through 2}
\label{Figure_3}
\end{figure}

Run-through 2 began with dancers starting the module {\em Clapping}. A snapshot at 38 seconds into Run-through 2 is shown in Fig.~\ref{Figure_3}(a).  In the snapshot, three dancers, Simon, Shayla, and Raja, can be seen performing {\em Clapping}. Raja had already begun modifying {\em Clapping} by clapping his hands on his knees. The module {\em Bumper Body} was introduced and the snapshot at 109 seconds in Fig.~\ref{Figure_3}(b) shows Cori, Chris, Natalie, and Tan, all performing {\em Bumper Body} while Shayla, Simon, and Raja continued in {\em Clapping}.  

When {\em Bumper Body} ended, {\em Folk} was introduced.  When {\em Folk} ended {\em Romper Room} was introduced and when {\em Romper Room} ended {\em Chris} was introduced.  All the while {\em Clapping} persisted.  And dancers frequently switched back and forth between {\em Clapping} and whichever other module was current.  For example, at 157 seconds in Fig.~\ref{Figure_3}(c), Natalie, Simon, Chris, and Shayla were all performing {\em Folk}.  But at 228 seconds in Fig.~\ref{Figure_3}(d), Shayla and Cori were the only two dancers performing {\em Folk}, and at 251 seconds in Fig.~\ref{Figure_3}(e), Raja was alone doing {\em Folk}.  Those dancers who had returned to {\em Clapping} were arranged in a line, and it was if the dancers were taking turns doing solo performances of {\em Folk}.

All of the dancers but Chris can be seen performing the module {\em Chris} at 368 seconds in Fig.~\ref{Figure_3}(f).  Chris was keeping {\em Clapping} alive.  In fact, {\em Clapping} had another resurgence of popularity before it  was abandoned and {\em Bravo} introduced.  At the nine minute mark, one dancer remained in {\em Chris} while the rest performed {\em Bravo}.

\subsection{Run-through 3}

In Run-through 3, the catalog of nine modules was the same as in Run-through 2.  However, the additional rule was changed: in Run-through 3, we asked the dancers to be more resistant to changing their choices, i.e., to {\em decrease} their tendency to switch modules.  The intent was to continue to examine switching tendency as a possible design dial. However, in contrast to Run-through 2 where the dial was turned up, in Run-through 3 the dial was turned down.

Run-through 3 began with the dancers starting the module {\em Clapping}. At 70 seconds into the run-through, the module {\em Box and Drag} was introduced. {\em Box and Drag} persisted while {\em Clapping} ended and {\em Kiss} was introduced, {\em Kiss} ended and {\em Folk} was introduced, {\em Folk} ended and {\em Bravo} was introduced.  {\em Box and Drag} was still active at the nine minute mark, along with {\em Bravo}.  

Throughout Run-through 3, there was minimal switching between active modules; for example, there was nothing like the switching of soloists performing {\em Folk} as seen in Run-through 2.  When Shayla introduced {\em Folk} at 286 seconds into Run-through 3, she performed it alone for 48 seconds before she was joined by Simon, and then some of the others.  At 450 seconds, Chris rose up from {\em Box and Drag} and caught each dancer who was performing {\em Folk} and folded the dancer  into {\em Box and Drag} until at 490 seconds, he managed to get everyone committed to {\em Box and Drag} and {\em Folk} was gone.  At just this moment, Chris introduced {\em Bravo}.  Later, after the nine minute mark, when almost all the dancers were performing {\em Bravo}, Cori seemed to play with the idea of folding some of them into {\em Box and Drag}, something like an echo of what we had seen Chris do earlier.

After Run-through 3, we ran one more run-through during which we tried out a global cue to signal a change in instruction, i.e., to signal a dial change. 
One beat on the drum directed the dancers to make more ``studied'' choices, whereas a scale on the xylophone directed the dancers to make more ``random'' choices.  The dancers were a bit worn out at this point, but they described enjoying the exercise. Raja described liking the connection to the external environment rather than only to internal, i.e., social, factors.  Cori reported liking the global external signal because it served to refresh in her mind a priority.

\subsection{Quantifying group dynamics}

\begin{figure}[t!]
\centering
\includegraphics[width=0.9\textwidth]{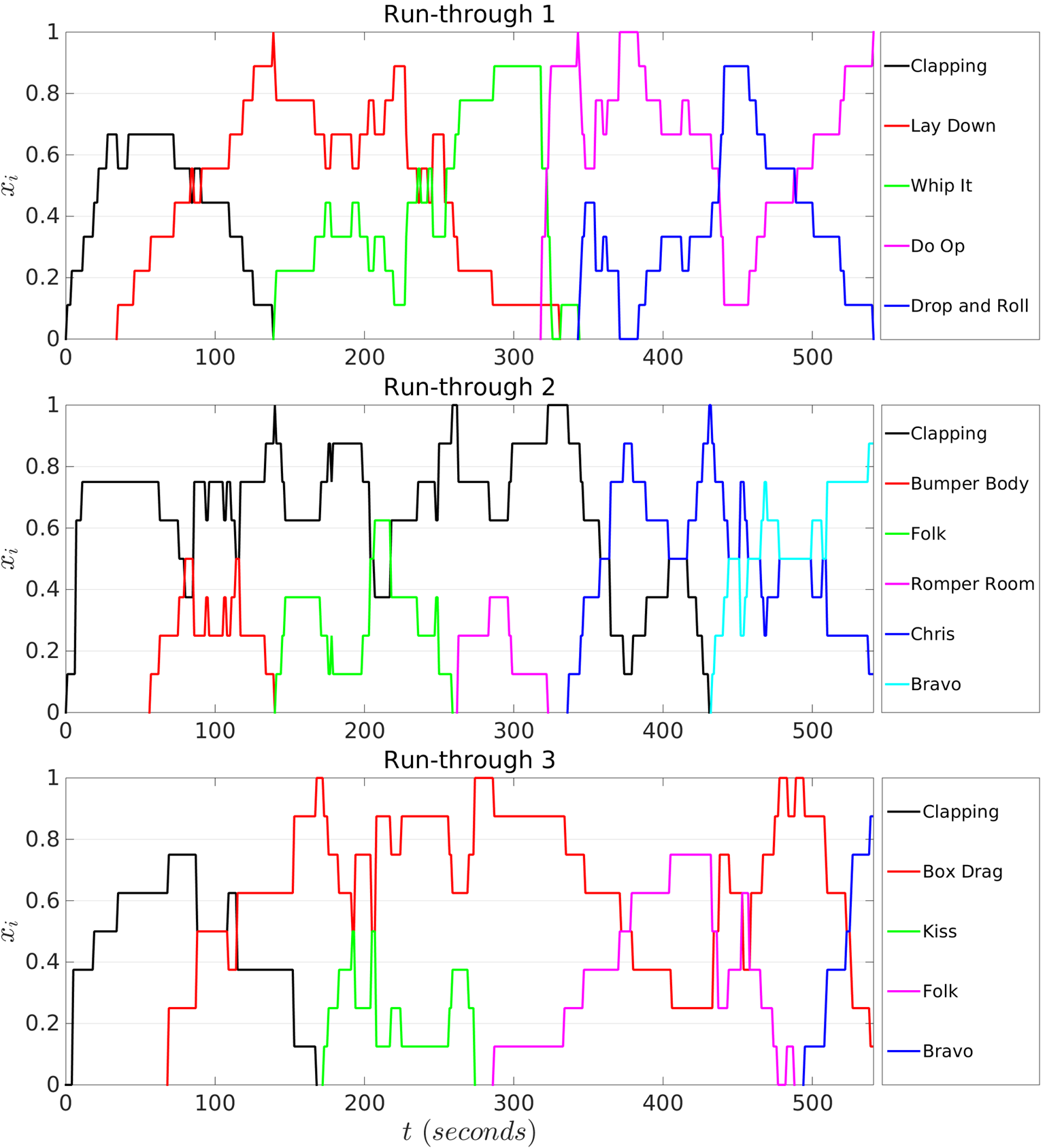}
\caption{The fraction of total number of dancers performing each of the at most two modules, $x_1$ and $x_2$ as a function of time $t$ during Run-through 1 (top plot), Run-through 2 (middle plot), and Run-through 3 (bottom plot). For Run-through 1, the $x_1$ and $x_2$ are fractions of a total of nine dancers performing each of the two active modules, whereas for Run-throughs 2 and 3, they are fractions of a total of eight dancers. The modules are distinguished by the colors indicated in the key on the right.}
\label{Figure_4}
\end{figure}

Motivated by our modeling approach, described below, we quantified the explore-exploit dynamics of the group of dancers in terms of the distribution of dancers over the modules.  For Run-throughs 1, 2, and 3, we visualized the changing distribution by plotting in Fig.~\ref{Figure_4} the fraction of total number of dancers performing each of the current modules as a function of time. Videos extracted from the cameras were synchronized using off-the-shelf video editing tools ({\em SI Appendix}, Fig. S1). The synchronized videos were then analyzed by counting the number of dancers for each one-second time period and the trajectories drawn by connecting the sample points. 

The top, middle, and bottom plots of Fig.~\ref{Figure_4}, show what fraction of the group of dancers in Run-through 1, 2, and 3, respectively, performed which modules, distinguished by color, over time.  Time, on the horizontal axis, is denoted by $t$ and expressed in seconds from the start of the run-through at $t=0$ seconds until the nine minute mark at $t=540$ seconds.  

We denote the fraction of dancers at time $t$ performing one of the two current modules by $x_1(t)$ and  the fraction of dancers at time $t$ performing the other of the two current modules by $x_2(t)$. The fractions $x_1(t)$ and $x_2(t)$ therefore take values between 0 and 1 for all $t$. Since only two modules were allowed at a time, $x_1(t) + x_2(t) = 1$ if all dancers were dancing at time $t$. For instance, at $t = 300$ seconds into Run-through 1, reading off the green line in the top plot of Fig.~\ref{Figure_4}, the fraction of dancers performing {\em Whip It} was $x_1(300) = 8/9$ (everyone but Rhonda) and reading off the red line, the fraction of dancers performing {\em Lay Down and Get Up} was $x_2(300) = 1/9$ (Rhonda).  

Typically there are two lines at any given time, one for each of the two active modules. When a module is completed, the fraction for that module goes to zero, e.g., the black line representing the fraction for {\em Clapping} at $t=146$ seconds in the top plot of Fig.~\ref{Figure_4}.  And when, a new module is introduced, a new line in a new color emerges, e.g., the green line representing {\em Whip It} at $t=146$ seconds in the top plot of of Fig.~\ref{Figure_4}.
%
%

%
%
\section{Evolutionary dynamic model}

Our investigations in the rehearsal described above were motivated by our analysis of a mathematical model that we derived using the replicator-mutator dynamics from evolutionary game theory \cite{Burger_98}. These dynamics describe a game played by a population of individuals, each of which has a fixed strategy, interacts randomly with others, and receives a reward, also called a payoff, which determines its success in the game. In terms of natural selection, payoff is interpreted as fitness and success as reproductive success: the higher the fitness associated with a strategy, the faster the strategy will reproduce \cite{Nowak2006}. 

Letting the interacting individuals in the game represent the interacting dancers, and the strategies represent the dance modules, the replicator-mutator dynamics provide a useful model of TMBO. One important reason is that the fitnesses, i.e., the rewards, in the replicator-mutator dynamics change over time; in fact, fitnesses depend explicitly on the changing distribution of the population over the different strategies, much like we expect artistic reward in TMBO to depend on the changing distribution of dancers over the active modules.   Another reason is that the rates of reproduction of strategies depend not only on fitness but also on mutation, which in the equations is given by a random term; this allows us to use mutation to represent uncertainty and spontaneity associated with the dancers' choices.  
Further, the outcomes of the replicator-mutator dynamics range from mixed strategy solutions, where multiple strategies coexist, to pure strategy solutions, where a single strategy is adopted by the entire population. This parallels our central interest: the tension between how dancers exploit active modules and how they converge on a single module so that a new one can be explored.

The outputs of the replicator-mutator dynamics are the time-varying fractions of the population associated with the different strategies. This corresponds nicely to the dance, where we seek to understand how the fraction of the population of dancers performing each active module changes as a function of time (as in Fig.~\ref{Figure_4}).  And, getting to the original motivation for our collaboration, the mathematical model offers a number of parameters that can be modulated to correspond to qualitative instructions, i.e., dials, that the dancers might receive either before or during a performance.

So we let the strategies in the replicator-mutator model represent the dance modules, and evolutionary time $t$ represent the time $t$ during the dance. 
We set the total number of strategies $N$ equal to the limit on number of dance modules that can be performed at a time. For our study we considered the replicator-mutator dynamics with $N=2$. We interpreted a drop close to zero in the fraction of one strategy followed by a rise in that fraction as the completion of one dance module followed by the introduction of a new module. In this way, the replicator-mutator dynamics could be used to represent the dancers' progression through some or all of the available catalog of modules, while respecting the rule limiting the number of  modules allowed to be performed at a time.

In the following we describe the model for the case of $N=2$. The more general model for $N \geq 2$ is provided in the {\em SI Appendix} S2. The replicator-mutator equations describe the changing fractions in a very large population, so we made the simplifying abstraction that the number of dancers in the group is very large. Then, the fraction of dancers $x_1(t)$ performing one of the active modules (call it module 1) at time $t$ and the fraction of dancers $x_2(t)$ performing the other of the active modules (call it module 2) at time $t$ can take any value in the interval from 0 to 1. Since $x_1(t) + x_2(t) = 1$ for all time $t$, given one fraction, e.g., $x_1(t)$, we can always find the other, as $x_2(t) = 1 - x_1(t)$. 

The time rate of change of $x_1$, denoted $\frac{d{x}_1}{dt}$, and the time rate of change of $x_2$, denoted $\frac{d{x}_2}{dt}$, are given by the replicator-mutator dynamics:
\begin{eqnarray}
\frac{d{x}_1}{dt} &= & (x_1 f_1 q_{11} + x_2 f_2 q_{21}) - \phi x_1 , \nonumber \\
\frac{d{x}_2}{dt} &= & (x_1 f_1 q_{12} + x_2 f_2 q_{22}) - \phi x_2.
\label{eqnrepmut}
\end{eqnarray}
Here, $x_1(t)$ and $x_2(t)$ are written as $x_1$ and $x_2$ for brevity.  And $x_1 f_1 q_{11}$ refers to the product (multiplication) of the three variables $x_1$, $f_1$, and $q_{11}$, and similarly for the other terms. The variables $f_1$ and $f_2$ are the fitnesses (payoffs or rewards) of module 1 and 2, respectively, and $\phi = f_1x_1 + f_2x_2$ is the average fitness over the two modules. $q_{21}$, respectively $q_{12}$, is the probability of mutating from module 2 to 1, respectively from module 1 to 2. $q_{11}$ and $q_{22}$ are the probabilities of not mutating. So the term in the first equation $x_1 f_1 q_{11} - \phi x_1$, which is equal to $x_1 \phi (q_{11}f_1/\phi  - 1)$, will contribute to growing $x_1$ if $q_{11} f_1/\phi  > 1$, i.e., if the relative fitness $f_1/\phi$ of module 1 is sufficiently large, or shrinking $x_1$ if $q_{11} f_1/\phi  < 1$, i.e., if the relative fitness $f_1/\phi$ of module 1 is sufficiently small. The other term $x_2f_2 q_{21}$ will modulate the growth or shrinkage of $x_1$ by a term in the rate that is proportional to $q_{21}$, the probability of mutating from module 2 to 1. 

Eq.~(\ref{eqnrepmut}) was originally derived to describe evolutionary biology. However, we were motivated in part to generalize it for our purposes because it can also be derived as the limit of a stochastic error-prone imitation process, where agents imitate a strategy, e.g., strategy $1$ with $x_1 > 0$, at a rate proportional to its relative fitness $f_1/\phi$ and mutate (e.g., because of an error) to strategy 1 from alternative strategy 2 at a rate proportional to $q_{21}$ \cite{Traulsen2006}.

In applications of the replicator-mutator dynamics, the fitnesses $f_1$ and $f_2$ are typically defined as {\em linear} functions of the fractions $x_1$ and $x_2$, as in \cite{NowakScience2001}. A linear fitness function implies that the sensitivity of fitness to changes in fractions is a constant. For example, if the fitness function is linear in $x_1$, then if $x_1$ is doubled, the fitness doubles no matter if $x_1$ is a small, intermediate, or large number.  We found that to provide a model rich enough to represent the dynamics of TMBO, we needed more subtle sensitivity of fitness to fractions. We thus generalized the fitness function so that it has low sensitivity to changes in $x_1$ and $x_2$ when they are very small (near 0) or very large (near 1) and high sensitivity to changes in $x_1$ and $x_2$ when $x_1$ and $x_2$ take intermediate values (near 1/2).  The idea was to make the response highly sensitive only for intermediate values of $x_1$ and $x_2$, when major qualitative transitions are prone to appear.

We defined the new nonlinear fitness functions using what's known as a Hill type function $\sigma_{\gamma,k}$ as follows (where $\gamma$ and $k$ are parameters):
\begin{eqnarray}
f_1= b_{11}\sigma_{\gamma,k}(x_1) + b_{12}\sigma_{\gamma,k}(x_2), \;\;\; \sigma_{\gamma,k}(x_1)=\frac{\left(\frac{x_1}{1-x_1}\right)^\gamma}{k+\left(\frac{x_1}{1-x_1}\right)^\gamma}, \nonumber \\
f_2 = b_{21}\sigma_{\gamma,k}(x_1) + b_{22}\sigma_{\gamma,k}(x_2), \;\;\; \sigma_{\gamma,k}(x_2)=\frac{\left(\frac{x_2}{1-x_2}\right)^\gamma}{k+\left(\frac{x_2}{1-x_2}\right)^\gamma}.
 \label{EQ:modified fitness}
 \end{eqnarray}
The coefficient $b_{11} \geq 0$ describes the strength of the dependence of $f_1$ on $x_1$ ($b_{12}$ the dependence of $f_1$ on $x_2$ and likewise for $b_{21}$ and $b_{22}$). For example, a large $b_{12}$ could represent the relative ease in transitioning commitment from module $2$ to module $1$. The function $\sigma_{\gamma,k}(x_1)$ is ``sigmoidal'' in $x_1$ (likewise $\sigma_{\gamma,k}(x_2)$ is sigmoidal in $x_2$), which means that it saturates at the value 1 for $x_1$ close to 1 and at the value 0 for $x_1$ close to 0. When $\gamma=k=1$, it specializes to the linear function $\sigma_{1,1}(x_1) = x_1$ (and $\sigma_{1,1}(x_2) = x_2$). The parameters $k$ and $\gamma$ determine the shape of the sigmoidal function $\sigma_{\gamma,k}$. Notably, for smaller values of $k$, $\sigma_{\gamma,k}$ has a steeper slope (becomes more sensitive) and saturates at small or large $x_1$ (see {\em SI Appendix}, Fig.~S2). 

The degree to which there is spontaneous or random switching between modules is governed by a probability $\mu$ called mutation strength, which takes a value in the interval from 0 to 1. We let 
\begin{equation}
q_{11}= q_{22} = 1-\mu, \qquad q_{12} =  q_{21} = \mu. 
\label{eq:Q}
\end{equation}
Then $\mu$ represents the probability of randomly switching from module 1 to 2 or 2 to 1, and ($1 - \mu$) represents the probability of not switching.

Given a dynamic system defined by nonlinear equations, a bifurcation analysis can be used to reveal the sensitivity of the dynamical behavior of the system to the value of a system parameter. A bifurcation refers to the change in number and stability of steady solutions of the dynamical equations as the parameter, called the bifurcation parameter, passes through a critical value, called the bifurcation point. Bifurcation analysis of the dynamics of equations (\ref{eqnrepmut}) - (\ref{eq:Q}), in the linear case $\gamma=k=1$, show the solutions and their stability to be sensitive to the value of bifurcation parameter $\mu$ \cite{NowakScience2001,DP_CHC_NEL,Levin_2010}.    

Consider the case of symmetric fitness coefficients  in which $b_{11} = b_{22} = 1$  and $b_{12} = b_{21} = b$ with $0<b<1$. If $\mu$ is large enough, there is a single stable distributed equilibrium (mixed strategy solution) corresponding to $x_1 = x_2 = 0.5$, i.e., the population is distributed equally over the two modules. If $\mu$ is small enough, the distributed equilibrium is unstable, and there are two stable solutions, each corresponding to one strategy dominating the other in the population.  In the limit as $\mu$ decreases to zero there is bi-stability of the two fully dominating module equilibria (pure strategy solutions) corresponding to $x_1 = 1$, $x_2 = 0$ and $x_1 = 0$, $x_2=1$ \cite{NowakScience2001}, i.e., the stable solutions correspond to the population fully committed to one or the other module. Thus, the magnitude of $\mu$   determines if the population is uniformly distributed over the modules or if a single module dominates.

As a model of TMBO we interpret the distributed equilibrium, $x_1 = x_2 = 0.5$ as full exploitation, since it reflects juxtaposition of current modules with half the group performing module 1 and half the group performing module 2. We interpret the fully dominating module equilibria as the prerequisites for exploration, since they reflect the opportunity to introduce one or more new modules with the majority of the group performing one of the modules and a minority performing the other. The bifurcation analysis suggested that $\mu$ is a key control in the explore-exploit tradeoff. Since the explore-exploit tradeoff in the dance was observed to be dynamic, we augmented our model with dynamics for $\mu$. That is, instead of letting $\mu$ be a fixed parameter, we defined feedback dynamics for $\mu$ that represent how the dancers might have been modulating something like $\mu$ based on their observations of how many dancers were performing each module ($x_1$ and $x_2$). Because dancers are trained to be highly physically aware, they will have good estimates, not only of the fractions $x_1$ and $x_2$, but also of the presence or absence of a dominating module, i.e., whether the group is fully exploiting or in a position to explore a new module.

The dynamics of $\mu$ that we introduced into the model (Eq.~(\ref{two_mod_mu}) below) allow flexibility in representing how the dancers react to the presence or absence of a dominating module, i.e., to the presence or absence of a large majority of the dancers performing one module. The intuition is that the model should allow for the dancers to cycle between the presence and absence of a dominating module, and in turn cycle between exploring and exploiting as artistic rewards for choices grow and decline over time.  So we designed the dynamics such that if no module is dominating, $\mu$ decreases and drives the fractions towards a dominating module. Likewise, if one of the modules is dominating, $\mu$ increases and drives the fractions away from the dominating module solution. How sensitive the reaction is depends on the remaining model parameters, notably, the sensitivity parameter $k$ in $\sigma_{\gamma,k}$.  Thus, $\mu$ and $k$, or some combination of the two, might be used to translate the notion of switchiness from the rehearsal to the model, i.e., modulation of $\mu$ and $k$ in the model might provide a useful representation of a switchiness dial for the dancers.

Let $K > 0$ be a time-scale parameter that regulates how fast $\mu$ changes relative to how fast $x_1$ and $x_2$ change.  Let $0< \alpha_1 < 0.5$ and $\alpha_2 = 1 - \alpha_1$ be thresholds.  We define the time rate of change of $\mu$, denoted $\frac{d\mu}{dt}$, as 
\begin{equation}
\frac{d\mu}{dt} = K (x_1 - \alpha_1)(x_1 - \alpha_2)\mu (1 - \mu). \\
\label{two_mod_mu}
\end{equation}
Since $x_2 = 1 - x_1$, the right side of Eq.~(\ref{two_mod_mu}) can equivalently be written as $K(x_2-\alpha_1)(x_2 - \alpha_2)\mu(1-\mu)$. As intended these dynamics imply that if the fractions are close to the distributed solution where $\alpha_1 < x_1 < \alpha_2$ and $\alpha_1 < x_2 < \alpha_2$, then one of the terms in the product on the right side of Eq.~(\ref{two_mod_mu}) will be negative so that $\mu$ will decrease and the fractions will move towards a dominating solution. Alternatively, if the fractions are close to a dominating solution where $0<x_1< \alpha_1$ and $\alpha_2 < x_2 < 1$ or $0<x_2< \alpha_1$ and $\alpha_2 < x_1 < 1$, then the product on the right side of Eq.~(\ref{two_mod_mu}) will be positive so that $\mu$ will increase and the fractions will move towards a distributed solution. See the {\em SI Appendix} S2 for a more general formulation of the dynamics of $\mu$. An earlier version of these dynamics is described in \cite{OzcimderACC2016}.  

We note that the dynamics of $\mu$ depend not only on $k$ but also on $\alpha_1$ and $\alpha_2$.  So $\alpha_1$ and $\alpha_2$ could be investigated as to their influence on the overall dynamics and thus also as possible representatives of  dials for the dancers.  Indeed, in the model of \cite{DeyACC2018} we extend the model described here by introducing dynamics for $\alpha_1$ and $\alpha_2$, which results in even more varied and rich dynamic behavior. 
%
%
%

%
%
\section{Results from the studio}
%

As discussed above, we quantified the explore-exploit dynamics of the group, in both human and modeled settings, in terms of the changing distribution of dancers over the modules.   The changing distribution can be visualized by plotting the fraction of total number of dancers performing each of the current modules as a function of time.  As seen in Fig.~\ref{Figure_4}, the top, middle, and bottom plots show the fractions $x_1$ and $x_2$ of the group of dancers in Run-through 1, 2 and 3, respectively, as a function of time.   

During the 540 seconds of Run-through 1, as can be observed in the top plot of Fig.~\ref{Figure_4}, the dancers performed five of the nine available modules in the catalog.  At all times except for the beginning when the dancers were first entering and during the brief period at $t= 318$ seconds, it can be observed that there were only two active modules. During the short interval at $t=318$ seconds, a third module ({\em Do Op}) was introduced a moment or two before the lingering dancer (Rhonda) had completed the second active module ({\em Lay Down and Get Up}). 

The top plot of Fig.~\ref{Figure_4}  shows that one after another the dancers started with {\em Clapping} until, at $t = 34$ seconds, there were six dancers in {\em Clapping} ($x_1 = 0.67$). One dancer then introduced {\em Lay Down and Get Up}, and others followed until at $t=57$ seconds all nine dancers were engaged in either {\em Clapping} or {\em Lay Down and Get Up}. From $t=70$ to 110 seconds the nine dancers were evenly distributed over the two modules, with some switching of membership. Then, $x_2$ grew while $x_1$ shrunk, until at $t=135$ second {\em Clapping} was gone ($x_1=0$) and {\em Lay Down and Get Up} was fully dominating ($x_2=1$). At that moment one and then another dancer introduced the new module {\em Whip It} and two others joined so that at $t=170$ second, $x_3 = .44$ and $x_2= .56$. This distributed state persisted for a while but then {\em Lay Down} dominated again ($x_2 = .89$). However, {\em Whip It} did not go away. Instead, a few seconds later, there was a return to the distributed state followed by a period in which {\em Whip It} dominated ($x_1 = .89$). At $t = 318$ sec {\em Do Op} was introduced and quickly dominated ($x_4 = 1$). A few second later {\em Drop and Roll} was introduced and there followed oscillations between a dominating module and a distributed state until at the end {\em Do Op} was fully dominating.

Over the 540 second shown in all three of run-throughs in Fig.~\ref{Figure_4}, the dancers can be observed to dynamically balance the explore-exploit tradeoff by oscillating as a group between a dominating module (prerequisite to exploration) and a distribution over two modules (exploitation). The frequency of the oscillations was irregular and the module that became dominant in each cycle was irregularly chosen. The order of introduction of modules was not necessarily the order of completion of modules. For example, in Run-through 1, {\em Lay Down} lasted as long as {\em Whip It} even though it was introduced well before and {\em Do Op}, introduced before {\em Drop and Roll}, outlasted {\em Drop and Roll}. In Run-through 2, {\em Clapping} outlasted {\em Bumper Body}, {\em Folk}, and {\em Romper Room}.  In Run-through 3, {\em Box Drag} was the second module introduced and it was still active at the end of the nine minute mark.

Although the explore-exploit tradeoff was not discussed with the dancers prior to the investigation in the studio, during the rehearsal the dancers described feeling a distinct tension between experimenting with current modules and moving on to introduce new modules. They also described finding new challenges and creative opportunities with the new rules, most especially the limit to two modules at a time. For example, dancers who resisted joining a popular module influenced the dynamics in new ways. This was very apparent when Rhonda resisted giving up on {\em Lay Down and Get Up} in Run-through 1. By exploiting the less-populated module, Rhonda actively blocked exploration. 

Likewise, dancers who worked to recruit others to their module invented new variation. By recruiting to the more-populated module, these dancers actively advanced exploration.  The snapshot in Fig.~\ref{Figure_2}(e) shows eight dancers in a circle around Rhonda, on the floor, clearly urging her through their physical orientation and facial expressions to join them in {\em Whip It}. In Run-through 3, Chris went up to several dancers performing {\em Folk} and folded them into {\em Box and Drag} until no one remained in {\em Folk}.  He then very quickly introduced {\em Bravo} before anyone could go back to {\em Folk}.  Later Cori created a moment evocative of Chris' strategy by folding dancers who were performing {\em Bravo}.  However, Cori did not appear to be insistent and it read as playful and humorous.  

The constraints imposed by the modified rules and heightened tension were seen to motivate new dramatic moments. And because the dancers were allowed to switch between active modules, rather than being required to follow the order that modules were introduced, they reported that they could hold on to an idea and find a time to switch back to a module to pursue their idea when they saw the chance.  As a result of the experience of the investigation in the studio, the new rules became an ongoing part of the practice and the language of explore-exploit became a baseline for development of the piece.  In performance, even when the limit of two modules was lifted and the other modifications removed, the dancers created moments inspired by these rules.

\begin{figure}[t!]
\centering
\includegraphics[height=4.2cm,width=12.375cm]{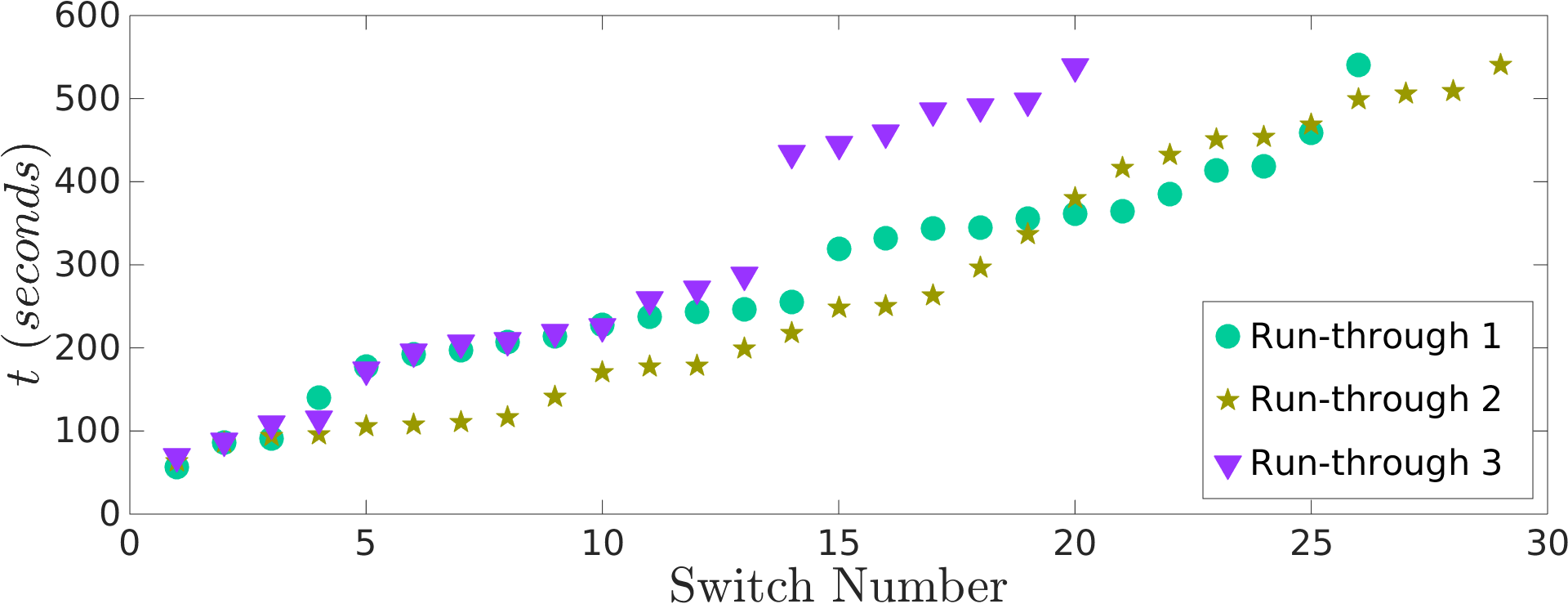}
\caption{Switching times for Run-throughs 1, 2, and 3. Each switching time refers to a time when the value of $x_1$ (and therefore $x_2$) changed from increasing to decreasing or vice versa.}
\label{Figure_5}
\end{figure}

Run-throughs 2 and 3 were also used to investigate the possibility of giving directions, like a design dial, to the dancers that would be analogous to manipulating design parameters in the model (like $\mu$ and $k$) to modify the collective behavior. This was investigated by asking the dancers to modify their switching tendency. The results suggested that switching tendency could be modified through instruction and would shape the dynamics in interesting ways. For example, the different instructions on switching tendency were reflected in the different total numbers of modules introduced in the first 480 seconds of the improvisations: five modules in Run-through 1, six in Run-through 2, and four in Run-through 3. The higher number of modules is consistent with impulsiveness, i.e., a dial up in switching tendency (the instruction  in Run-through 2) and the lower number with resistance, i.e., a dial down in switching tendency (the instruction in Run-through 3). A similar effect can be observed in Fig.~\ref{Figure_5}, which shows all of the times at which $x_1$ (and therefore $x_2$) switched from increasing to decreasing or vice versa for each of the three run-throughs. These switches were most frequent and numerous for Run-through 2, when switching tendency was high, which yielded a total of 29 switches. They were least frequent and numerous for Run-through 3, when switching tendency was low, which yielded a total of 20 switches. We note, however, that this analysis might even underestimate the effect of the switching tendency instruction, since it does not account for switches by dancers that didn't change the overall fractions.  For example, in Run-through 2, dancers did a lot of switching between modules, e.g., between {\em Clapping} and {\em Folk}.  But often one dancer in {\em Folk} would return to {\em Clapping} just as another dancer in {\em Clapping} took their place in {\em Clapping}, thus leaving the overall fractions unchanged.
%
%
%

%
%
\section{Results from the model}
%
A bifurcation analysis of the model was used to investigate the roles of $\mu$ and $k$ in the modeled behavior of the group. The analysis influenced, and was influenced by, the investigation in rehearsal. To visualize the results of the analysis, we used a bifurcation diagram, which shows how steady solutions of the model dynamics change for different values of the bifurcation parameter $\mu$, which in turn depends on $k$.  We considered the symmetric case in which $b_{11} = b_{22} = 1$, $b_{12} = b_{21} = b$. The steady solution is represented by either $x_1$ or $x_2$, and, since $x_1+ x_2=1$, if one steady solution is $x_i$, then the other steady solution is $1-x_i$.  So the bifurcation diagram is a plot of steady solutions $x_i$ (think of it as $x_1$) on the vertical axis versus bifurcation parameter $\mu$ on the horizontal axis.

\begin{figure}[b!]
\centering
\includegraphics[height=12.4cm,width=10.0cm]{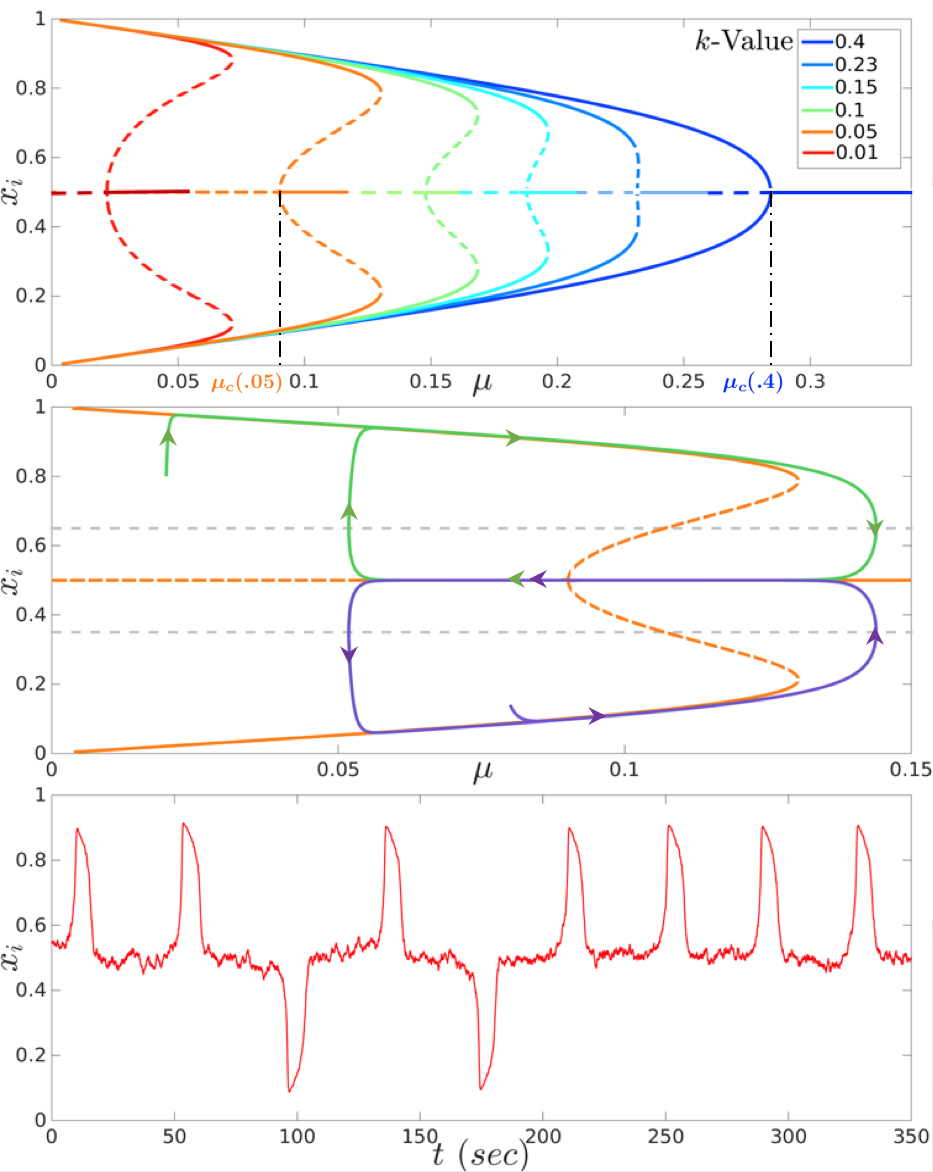}
\caption{Evolutionary model with $N=2$, $b=0.04$, $\gamma=2.5$. (Top) Bifurcation diagrams of Eqs.~(\ref{eqnrepmut})-(\ref{eq:Q}) showing steady solutions $x_i$ versus $\mu$ for six different values of $k$ in the nonlinear fitness function. Solid lines are stable solutions and dashed lines are unstable solutions. (Middle) Bifurcation diagram from top panel for $k=0.05$ with green and purple curves, each showing path of solution $x_i$, for a different initial condition, and for increasing and then decreasing $\mu$. (Bottom) $x_i$ as a function of time for Eqs.~(\ref{eqnrepmut})-(\ref{two_mod_mu}) with $K=0.05$, $\alpha_1 = 0.25$, $\alpha_2=0.75$, and noise variance $S(\cdot)$ a Gaussian with mean 0.5 and standard deviation 0.05. }
\label{Figure_6}
\end{figure}

Six distinct bifurcation diagrams for equations (\ref{eqnrepmut})-(\ref{eq:Q}) with $b = 0.04$, are plotted in the top panel of Fig.~\ref{Figure_6}. These plots show how the steady solutions $x_1$ (and $x_2 = 1 - x_1$) and their stability vary over a range of values of  $\mu$ for six different values of $k$ (see {\em SI Appendix S4} for a proof). Each of the six bifurcation diagrams is plotted in a different color as indicated by the key. Solid lines indicate solutions that are stable and dashed lines indicate solutions that are unstable. If the solution is stable, then it remains steady even in the case of a small change in conditions.  If the solution is unstable, however, even the smallest change in conditions will drive the dynamics away from the unstable solution and towards a stable solution. 

For every value of $\mu$ and $k$, there is a steady solution at the distributed solution $x_i = 0.5$, corresponding to half the population in module 1 and half in module 2.  This means that all six bifurcation diagrams have horizontal lines at $x_i=0.5$. To avoid overlapping lines from the different diagrams,  we only show the solution $x_i= 0.5$, for each of the six bifurcation diagrams, along the line near its intersection with the curved part of the diagram.  This point of intersection is known as the pitchfork bifurcation point.  We denote by $\mu_c(k)$ the value of $\mu$ at the pitchfork bifurcation point; we write $\mu_c$ as a function of $k$ because the location of the pitchfork bifurcation point depends on the value of $k$. In Fig.~\ref{Figure_6}, we identify $\mu_c$ for $k=0.05$ in orange and $\mu_c$ for $k=0.4$ in darkest blue. In each bifurcation diagram, the steady solution $x_i=0.5$ is stable for $\mu > \mu_c$ (solid line) and unstable for $\mu < \mu_c$ (dashed line).  Thus, in Fig.~\ref{Figure_6} every solid line on $x_i = 0.5$ continues to the right (as $\mu$ increases) and every dashed line on $x_i = 0.5$  continues to the left (as $\mu$ decreases).

When $k=0.4$ the bifurcation diagram (darkest blue) reveals a ``supercritical'' pitchfork bifurcation, represented by the c-shaped curve that opens to the left, at bifurcation point $\mu = \mu_c(0.4) \approx 0.28$. The diagram shows that for $\mu > \mu_c$ the distributed solution $x_i = 0.5$ is the only (stable) solution. For $\mu < \mu_c$ the distributed solution is unstable and the two (symmetric) dominant solutions are stable: $x_1 > 0.5$ ($x_2 < 0.5$) and $x_1 < 0.5$ ($x_2 > 0.5$). The magnitude of the stable dominant solutions increases smoothly as $\mu$ decreases. Thus, were $\mu$ to slowly decrease (increase)  past the bifurcation point, the result would be a slow change from the distributed solution to a dominant solution (vice versa).

The results from the studio in Figs.~\ref{Figure_4} and~\ref{Figure_5} show instead relatively rapid switching in $x_i$, corresponding to oscillations in exploration versus exploitation. This suggests that the model with $k\geq 0.4$, which includes the linear fitness model from the literature, is insufficient to fully capture the observed behavior.  However,  in our model with nonlinear fitness function, rapid switching does follow for slowly varying $\mu$ if $k$ is sufficiently small, much like the appearance of oscillatory behavior in relaxation oscillators (see {\em SI Appendix} S5). This can be understood from the bifurcation diagrams in the top plot of Fig.~\ref{Figure_6} where $k < 0.4$.  

For example, when $k=0.05$ the bifurcation diagram (orange) reveals a ``subcritical'' pitchfork bifurcation, distinguished by the c-shaped curve that opens to the right, at bifurcation point $\mu_c = \mu_c(0.05) \approx 0.09$. As in the supercritical case, when $\mu < \mu_c$ the distributed solution is unstable and the two (symmetric) dominant solutions are stable.  However, the situation is a little more complicated when $\mu > \mu_c$.   Here, before $\mu$ gets very large, there are fives steady solutions, three stable (solid) and two unstable (dashed). It is this more complicated diagram that leads to the relatively rapid switching in the fractions $x_1$ and $x_2$ as $\mu$ changes.

We illustrate this with the green and purple curves in the middle plot of Fig.~\ref{Figure_6}, which are model simulations plotted on top of the bifurcation diagram in the case $k=0.05$.  The simulated curves show how the solution $x_1$ changes for two different initial conditions as $\mu$ is slowly increased and then slowly decreased past the subcritical bifurcation point. The initial conditions for the green curve were approximately $x_1 = 0.8$ and $\mu = 0.02$, and $x_1$ can be seen first to grow closer to 1 then to shrink to 0.5 and then to grow back towards 1. The initial conditions for the purple curve were approximately $x_1 = 0.1$ and $\mu = 0.08$, and $x_1$ can be seen first to shrink closer to 0 then to grow to 0.5 and then to shrink back towards 0. Each loop represents a cycle between exploration and exploitation in the modeled dynamics.  In the green curve, it is module 1 that the population converges on before exploring a new module, whereas in the purple curve it is module 2.

The corresponding oscillatory behavior of $x_1$ as a function of time $t$ is represented in the bottom plot of Fig.~\ref{Figure_6}: one circuit around the green periodic solution in the middle plot corresponds to one up-down oscillation in the bottom plot and one circuit around the purple periodic solution in the middle plot corresponds to one down-up oscillation in the bottom plot.

The plot of $x_1$ versus $t$ in the bottom panel of Fig.~\ref{Figure_6} is a simulation of Eqs.~(\ref{eqnrepmut})-(\ref{two_mod_mu}) for $N=2$, $b=0.04$, $k=0.05$, $\gamma= 2.5$, $K=0.05$, $\alpha_1 = 0.25$, $\alpha_2=0.75$. Here $\mu$ is not independently driven up and down, but rather it follows its own dynamics as prescribed by Eq.~(\ref{two_mod_mu}). That is, Eqs.~(\ref{eqnrepmut})-(\ref{two_mod_mu}) are the dynamics of the distribution of the population between the two active modules with small $k$ and feedback dynamics for $\mu$ corresponding to more switching when one module is dominant and less switching when no module is dominant. These dynamics yield relatively rapid oscillations as observed in the rehearsal. 

In this simulation, white noise is added to Eq.~(\ref{eqnrepmut}) with variance $S(x_i)$ that depends on $x_i$ to represent uncertainties (like that observed in Fig.~\ref{Figure_4} at $t=318$ sec). $S(\cdot)$ is designed to be symmetric about $x= 0.5$, vanishing on the boundaries, i.e. $S(0)=S(1)=0$, and greatest at $x_i=0.5$, i.e., when it is more difficult for dancers to judge distribution across modules. The noisy response around $x_i = 0.5$ can be observed in the simulation of Fig.~\ref{Figure_6}; it leads to switching between the two different periodic solutions (green and purple in the middle plot).

We note further, from the top of Fig.~\ref{Figure_6}, that as $k$ increases the magnitude of oscillations decreases, and oscillations exist for higher values, and a smaller range, of $\mu$.   The insight here is that a small change in $k$ can shape the oscillations in the explore-exploit dynamics of the model. 

\begin{figure}[t!]
\centering
\includegraphics[height=7.875cm,width=12.0cm]{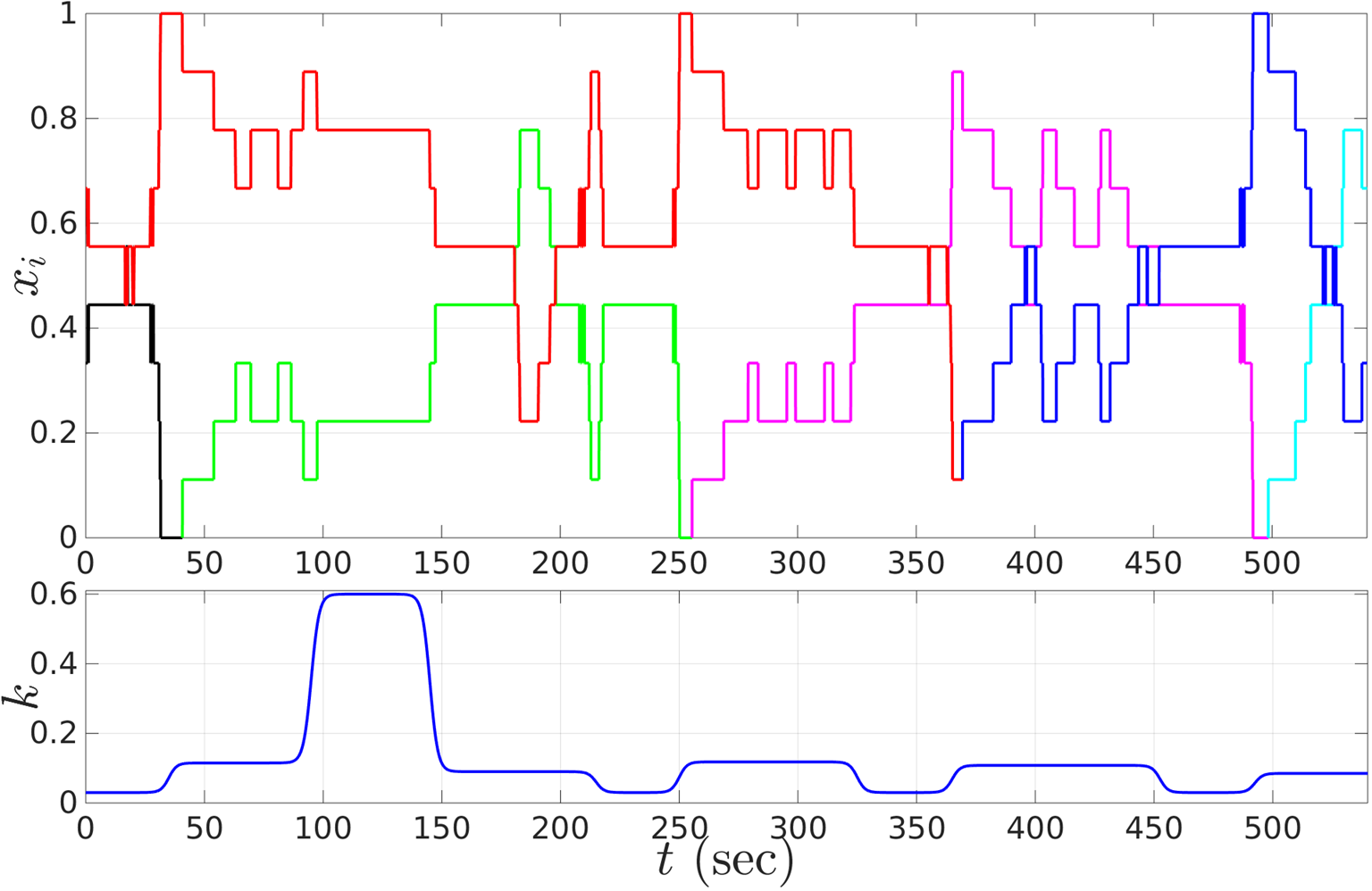}
\caption{Model simulation results for comparison with results in rehearsal plotted in Fig.~\ref{Figure_4}. (Top) The fraction $x_i$ of a population of nine as a function of time $t$ for a simulation of the model with $N=2$, $b=0.04$, $\gamma=2.5$, $K=0.05$, $\alpha_1 = 0.25$, $\alpha_2=0.75$, $S(\cdot)$ a Gaussian with mean 0.5 and standard deviation 0.05, and time-varying $k$. When $x_i$ goes to zero, a new module is introduced. Different modules are distinguished by different colors. (Bottom) Parameter $k$ varies with time $t$.}
\label{Figure_7}
\end{figure}

To make a qualitative comparison of the model output with observations from the studio (Fig.~\ref{Figure_4}), we plot in Fig.~\ref{Figure_7} a model simulation of $x_1$ and $x_2$ as a function of time $t$, discretized for a population of nine. Only two modules are allowed at a time and once $x_1$ or $x_2$ goes to zero, the corresponding module is replaced by a new module indicated by a new color. The simulation uses the same parameters as in Fig.~\ref{Figure_6}, except that $k$ is varied over time, as shown in the bottom panel of Fig.~\ref{Figure_7}, to represent variability in sensitivity over time. 

The simulation in Fig.~\ref{Figure_7} illustrates the influence of $\mu$ and $k$ on the explore-exploit tradeoff; 
it can be seen that changes in these parameters affected the model dynamics much the same way as the change in instruction in Run-throughs 2 and 3 (for more and less tendency to switch) affected the collective behavior as seen in the middle and bottom plots of Fig.~\ref{Figure_4}. For example, during the period from $t=100$ to 140 sec in the simulation of Fig.~\ref{Figure_7}, $k$ was dramatically increased resulting in a pause in oscillations in $x_1$ and $x_2$. We observe in Fig.~\ref{Figure_7} features in common with all of the run-throughs in Figs.~\ref{Figure_4}: oscillations with irregular frequency, irregular choice of dominating module in each cycle, and different ordering of introduced versus completed modules.
%
%
\section{Conclusions and future directions}
%

Until we began our experiments together, TMBO existed as a dance without music, imposing the general rule of staying within three modules of one another and not allowing skipping or switching between modules. Following the underlying principle that the dance should be different every time and the dancers should be looking for something different in the moment, we employed new approaches to examine how the rules and constraints could be modified and designed to support creativity and bring forth new experiences. By tuning our collective attention to understanding the explore versus exploit tension, where spontaneity and risk-taking could be artistically rewarding, we widened the dramatic possibilities of the score and heightened  the performers' knowledge of how they prefer to engage within the group, which then created new possibilities for both openness and control. We permanently adapted our rules to include sections where external mechanisms signal a time to adapt responsive behaviors, i.e., to turn the ``dials'' in one direction or another.     In the year following our joint work, Dan Trueman completed the composition of forty-four music modules in collaboration with the music ensembles So Percussion and Mobius Percussion, and we produced a fully realized version of TMBO with 15 dancers and 12 percussionists at New York Live Arts in March 2016. TMBO has since been reprised on several occasions, including at The Scotia Festival of Music, Nova Scotia, Canada, in June 2017 (for a video of the performance see \cite{NSTMBO2017}). 

In becoming an integral part of the development of the dance, our art and science investigation of TMBO produced new perspectives on social decision-making dynamics and opportunities for design in a real-time, collaborative, art-making context.  By integrating studies with the TMBO dancers in rehearsal and analysis of a representative mathematical model, we focused in on a simply parameterized mechanism to help 
describe and shape how dancers in TMBO address an explore-exploit tension that derives from competing artistic goals and constraints imposed by the choreographic rules. 

In most contexts studied in the explore-exploit literature, rewards associated with decision-making options do not change with time.  Decision-makers explore and exploit until they find the best option, and then they stop exploring. In TMBO, however, the rewards continue to change and so the performers continue to explore. An active module that is appealing or provocative at one moment will eventually become less so with time or with changing circumstances. A performer may first exploit the active module and then switch to exploring by introducing or joining in on a new module.  The new module may in turn lead others to exploit. By switching between exploring and exploiting, the dancers keep fresh the sequencing, timing, counterpoint, relationship, and variation of modules.
The dancers' switching behavior was observed in the oscillatory dynamics of the fractions $x_1$ and $x_2$ of the total number of dancers committed to each of the two active modules (Fig.~\ref{Figure_4}).  

The oscillatory dynamics in $x_1$ and $x_2$ were likewise captured with a mathematical model, generalized from the replicator-mutator equations of evolutionary biology to include a nonlinear fitness function with sensitivity parameter $k$ and self-contained feedback dynamics for mutation rate $\mu$. For small enough $k$, the model reveals a symmetric subcritical pitchfork bifurcation, which provides a mechanism for rapid switching between exploration and exploitation as $\mu$ evolves. Modulation of the parameters $\mu$ and $k$ in the model appeared to account well for the subtlety and variability of the dancers' innovative impulses in reactions to what the others were doing, thus providing a possible representation of instruction dials that could be used to signal rule changes to the dancers.  For example, changes in $k$, which in turn changed the dynamics of $\mu$, were shown to affect the frequency of the explore-exploit oscillations (Fig.~\ref{Figure_7}), much like what was observed in rehearsal (Fig.~\ref{Figure_4}) when dancers were asked to modulate their switchiness tendency.
In this case, a simple implementation of such a dial 
would be to define one beat of a drum as an instruction to the dancers to increase their switchiness tendency and two beats of the drum as an instruction to the dancers to decrease their switchiness tendency. 

The model can be used further as an off-line choreographic tool to examine the predicted consequences on collective explore-exploit dynamics of other kinds of design modifications. For example, study of the influence of asymmetries in the relative ease in transitioning from one module to another can be made using the parameters $b_{ij}$ \cite{Levin_2010}.  And study of the influence of thresholds in the shifting dynamics of mutation rate can be made using the parameters $\alpha_1$ and $\alpha_2$ \cite{DeyACC2018}. The case in which the limit on the number of current modules is greater than two can be studied using the replicator-mutator dynamics with $N > 2$ \cite{DP_CHC_NEL}. 

Our investigation, insights, and mathematical model may generalize to other contexts beyond structured improvisational dance, in the same way that the replicator-mutator dynamics have been applied in various contexts. For example, the simply parameterized but rich family of collective dynamics  may provide a useful framework for examining mechanisms of social decision-making in settings where expert performers together face an explore-exploit tension in which rewards change with time or with changing circumstances, such as in other art forms, collaborative design, or team sports.  The generalized model advances a related goal, which is to draw inspiration from the dynamics of natural systems and translate the mechanisms uncovered, using mathematics, to models that can be used for design \cite{GrayTCNS2018}.  In all these ways, our collaboration has proved to be a generative meeting of art and science.

\section*{Acknowledgments}
This research was supported in part by a grant for Artists and Scientists or Engineers from the Office of the Dean for Research at Princeton University.  The authors thank everyone who contributed to making TMBO, most especially the dancers.

\section*{Author contributions} K.O. and N.E.L. designed the research. K.O., R.L., D.T., and N.E.L. performed the research. K.O., B.D., A.F., and N.E.L. contributed new analytic tools. K.O., B.D., A.F., and N.E.L. analyzed the data. N.E.L. wrote the paper with input from all co-authors.

%

\newpage

\section*{Supplemental Information}

\renewcommand{\thesection}{S.\arabic{section}}
\renewcommand{\thesubsection}{\thesection.\arabic{subsection}}

\makeatletter
\def\tagform@#1{\maketag@@@{(S\ignorespaces#1\unskip\@@italiccorr)}}
\makeatother

\makeatletter
\makeatletter\renewcommand{\fnum@figure}
{\figurename~S\thefigure}
\makeatother

\setcounter{section}{0}
\setcounter{figure}{0}

\section{Synchronized video frames}

\begin{figure}[h]
\begin{center}
\includegraphics[width=0.90\textwidth]{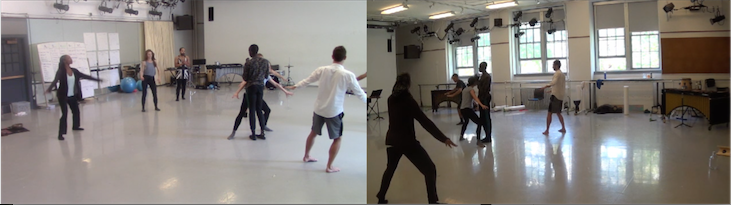}
\end{center}
\caption{Snapshots at the same time from each of the two synchronized videos of the dancers during Run-through 1.}
\label{Figure_S1}
\end{figure}

\section{General model for $N \geq 2$}
In the main text we specialize the model to $N=2$.  Here we present here the general model for the case that the limit on number of modules at a time is $N \geq 2$. We make the simplifying abstraction that the number of dancers in the group is very large so that we can represent the fraction of dancers $x_i(t)$ committed at time $t$ to strategy $i$ as a number in the interval $[0,1]$ for $i = 1, \ldots, N$. By definition $\sum_{i=1}^N x_i(t) = 1$ for all time $t$.  
The time rate of change of $x_i$ is given by the replicator-mutator dynamics:
\begin{equation}
\frac{d{x}_i}{dt} = \sum_{j=1}^N x_j f_j q_{ji} - \phi x_i ,
\label{eqnrepmut}
\end{equation}
where $f_i$ is the fitness of strategy $i$, $q_{ji}$ is the probability of mutating from strategy $j$ to $i$, and $\phi = \sum_{i=1}^N f_ix_i$ is the average fitness over the $N$ strategies. 

We define a new nonlinear fitness function $f_i$ using a Hill type function $\sigma_{\gamma,k}(x_j)$ as follows:
\begin{equation}
f_i=\sum_{j=1}^N b_{ij}\sigma_{\gamma,k}(x_j), \;\;\; \sigma_{\gamma,k}(x_j)=\frac{\left(\frac{x_j}{1-x_j}\right)^\gamma}{k+\left(\frac{x_j}{1-x_j}\right)^\gamma}.
 \label{EQ:modified fitness}
 \end{equation}
The coefficients $b_{ij} \geq 0$ describe the strength of the dependence of $f_i$ on $x_j$. 
The function $\sigma_{\gamma,k}(x_j)$ is sigmoidal in $x_j$ and specializes to the linear function $\sigma_{1,1}(x_j) = x_j$ for $\gamma=k=1$. 

Spontaneous switching between modules is governed by a probability $\mu \in [0,1]$ called mutation strength. Following \cite{DP_CHC_NEL}, we let 
\begin{equation}
q_{ii}=1-\mu \;\; {\rm for \; all} \; i, \qquad q_{ij} = \frac{\mu b_{ij}}{\sum_{l \neq i} b_{il}} \;\; {\rm for} \; \; i \neq j. 
\label{eq:Q}
\end{equation}
Then the probability that dancers spontaneously switch from module $i$ to any other module is represented by $\sum_{j=1,j\neq i}^N q_{ij} =\mu$.

In the linear, symmetric case in which $b_{ii} = 1$ for all $i$ and $b_{ij} = b \in (0,1)$ for $i \neq j$, if $\mu$ is large enough there is a single stable distributed equilibrium corresponding to $x_i = 1/N$ for all $i$, i.e., the population is distributed equally over all strategies. If $\mu$ is small enough the distributed equilibrium is unstable, and the stable solutions correspond to one dominant strategy.  In the limit as $\mu$ goes to zero there is multi-stability of a fully dominating strategy equilibria corresponding to $x_i = 1$, $x_j = 0$, $j \neq i$, i.e., the population is fully committed to one strategy. 

Let $K > 0$ be a time-scale parameter and $\alpha$ a real-valued function of $x_i$. To define dynamics for $\mu$, we introduce the scalar time-dependent variables $w_i(t)$ for $i=1, \ldots, N$ and let $\mu(t)$ depend on the
$w_i(t)$. The time rate of change of $w_i$ is
\begin{equation}
\frac{d w_i}{dt} = K \alpha(x_i) w_i (1 - w_i), \;\;\; w_i(0) \in [0,1]. \nonumber
\label{general_mu}
\end{equation}
So $w_i(t) \in [0,1]$ for all $t \geq 0$ and equilibria are $w_i=0$ and $w_i=1$. We can design the function $\alpha$ using thresholds on the $x_i$ to model how the dancers react to the presence or absence of a dominating module, i.e., to the presence or absence of a large majority of the dancers performing one module. For example, reinforcement of a dominant module can be modeled with $\mu = \min_i w_i$ and $\alpha(x_i) = \bar \alpha - x_i$ for $\bar\alpha \in (0.5,1)$. Then for any $i$ such that $x_i > \bar\alpha$, $w_i$ and thus $\mu$ will decrease, driving the state dynamics toward a dominating module equilibrium. If $x_i \leq \bar\alpha$ for all $i$, then all the $w_i$ and thus $\mu$ will increase, driving the state dynamics toward the distributed solution.

Alternatively, we can design the function $\alpha$ to model the opposite reaction of the dancers to the presence or absence of a dominating module.  We do this in the main text for $N=2$ in Eq. (4).

%
\section{Nonlinear fitness function}

The nonlinear fitness function given by Eq.~(1) in the main text depends on the sigmoidal function $\sigma_{\gamma,k}(x_j)$.  The parameters $\gamma$ and $k$ determine the shape of $\sigma_{\gamma,k}(x_j)$.  We illustrate the influence of $k$ on $\sigma_{\gamma,k}$ in Fig.~S2. 
\begin{figure}[H]  
\begin{center}
\includegraphics[scale=0.25]{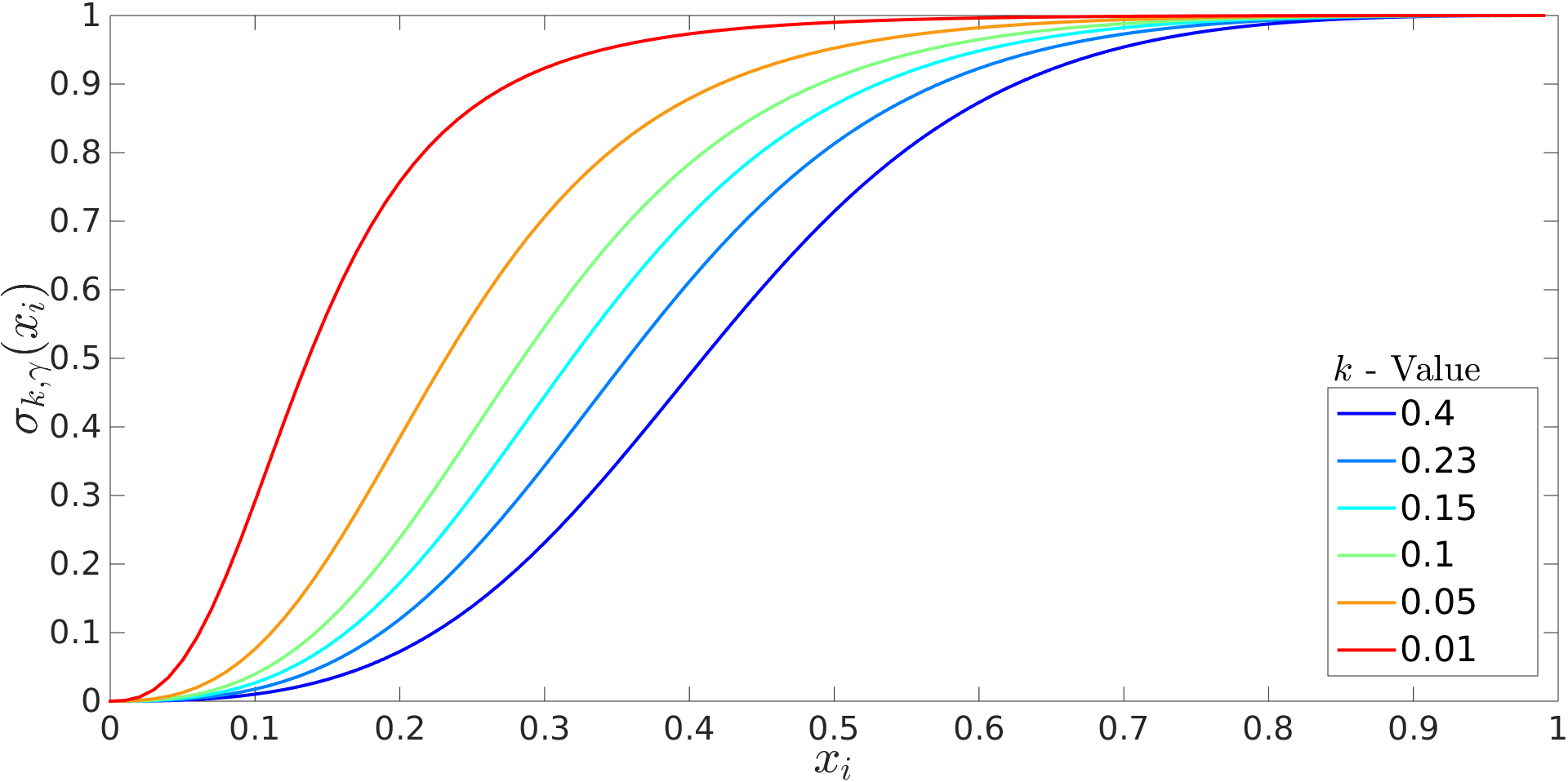}
\caption{The sigmoidal function $\sigma_{\gamma,k}(x_j)$ for different values of $k$ and $\gamma = 2.5$.}  
\end{center}
\label{Figure_S2}
\end{figure}


\section{Proof of symmetric quintic bifurcation}
\label{proof}

The proof of the supercritical pitchfork bifurcation is well known for the replicator mutator equations Eqs.~(1)-(3) from the main text with $N=2$ in the case of linear fitness function corresponding to $k=1$. The transition from the supercritical to the subcritical pitchfork illustrated in Fig.~6 is through a symmetric quintic pitchfork bifurcation with normal form $-x^5+\mu x$ \cite[Section~VI.5 and Figure~VI.6.1 top]{Golubitsky1985}.
We prove the existence of a symmetric quintic pitchfork bifurcation for the replicator mutator equations Eqs.~(1)-(3) from the main text with $N=2$ in the case of the new generalized nonlinear fitness function corresponding to small $k$.  In the case of $N=2$, $b_{11} = b_{22} = 1$, and $b_{12} = b_{21} = b \in (0,1)$, Eqs.~(1)-(3) specialize to 
\begin{eqnarray}
\dot x_1&=&g(x_1,\mu,k,\gamma,b) \nonumber\\
&:=&x_1(\sigma_{\gamma,k}(x_1)+b\sigma_{\gamma,k}(1-x_1))(1-\mu)+(1-x_1)(b\sigma_{\gamma,k}(x_1)+\sigma_{\gamma,k}(1-x_1))\mu-\phi x_1, \label{EQ:decoupled generalized}
\end{eqnarray}
where
\begin{eqnarray*}\label{EQ:modified fitness}
f_i=\sum_{j=1}^N b_{ij}\sigma_{\gamma,k}(x_j),
\end{eqnarray*}
and
\begin{eqnarray*}\label{EQ:modified fitness sigmoid}
\sigma_{\gamma,k}(x_j)=\frac{\left(\frac{x_j}{1-x_j}\right)^\gamma}{k+\left(\frac{x_j}{1-x_j}\right)^\gamma}.
\end{eqnarray*}
Our goal is to detect a symmetric quintic pitchfork in the scalar equation
$$g(x_1,\mu,k,\gamma,b)=0.$$
Since the system is $Z_2$ symmetric with respect to $x_1=0.5$ (that is, the change of variables $x_1\mapsto 1-x_1$ leaves Eq.~(S\ref{EQ:decoupled generalized}) invariant), this singularity has codimension 1 (\cite{Golubitsky1985}, Theorem VI.5.1(3)).  Thus, we only need one unfolding parameter.  For analytical tractability and in line with the main text, we fix $b=0.04$, $\gamma=2.5$ and use $k$ as the unfolding parameter.

\begin{figure}[h]
\centering
\includegraphics[width=0.5\textwidth]{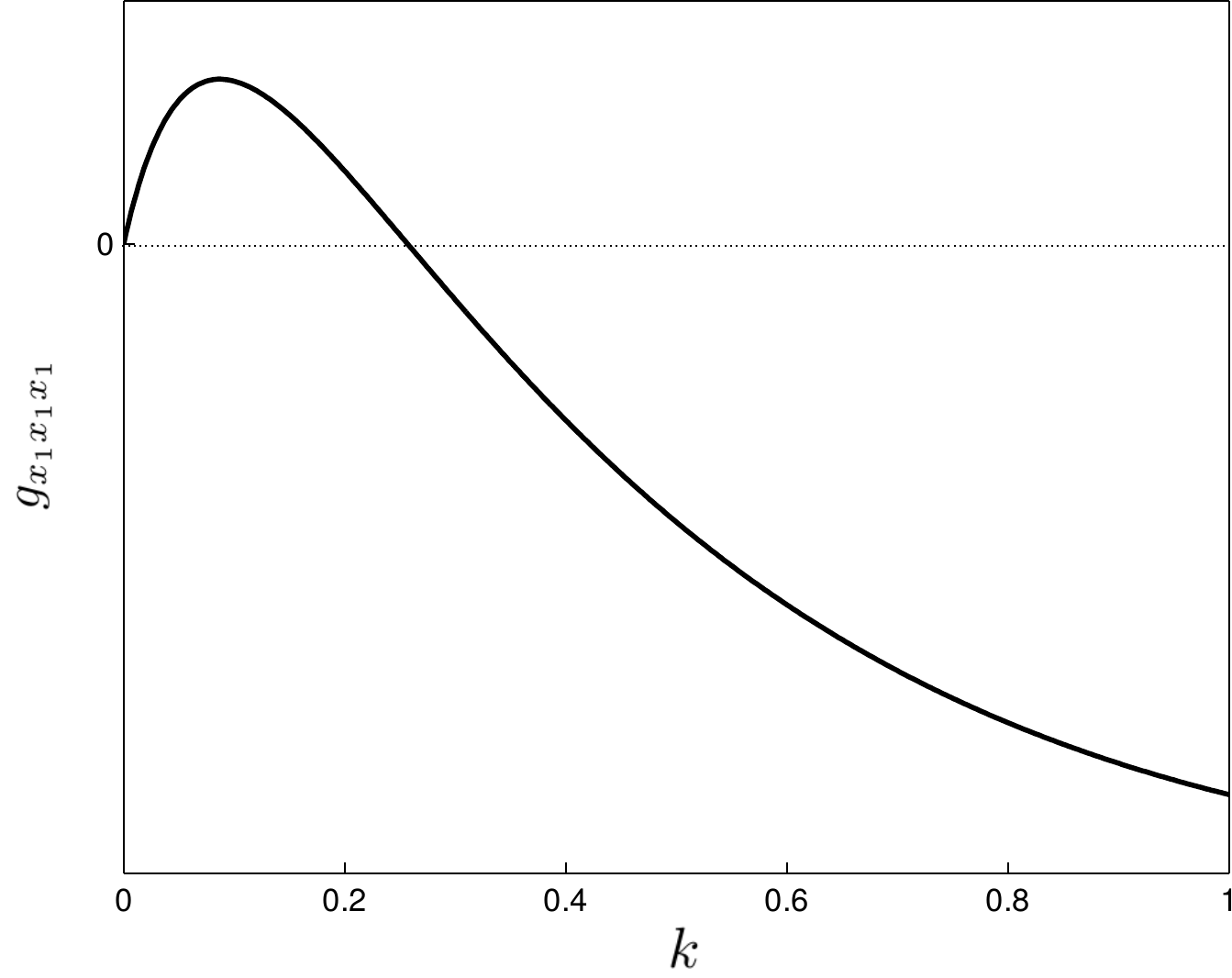}
\caption{Evolution of $g_{x_1x_1x_1}$ as a function of $k$.}
\label{Figure_S3}
\end{figure}

Following the recognition problem (\cite{Golubitsky1985}, Table VI.5.3), we seek $\mu^*,k^*$ such that
$$g(0.5,\mu^*,k^*,2.5,0.04)=g_{x_1}(0.5,\mu^*,k^*,2.5,0.04)=g_{x_1x_1x_1}(0.5,\mu^*,k^*,2.5,0.04)=0, $$
$$ g_{x_1x_1x_1x_1x_1}(0.5,\mu^*,k^*,2.5,0.04)\neq0\neq g_{x_1\mu}(0.5,\mu^*,k^*,2.5,0.04).$$
We first solve $g=0$ in terms of $\mu$ and plug the solution into the equation $g_{x_1x_1x_1}=0$ so that it becomes a function only of $k$. The equation $g_{x_1}=0$ is automatically solved by picking $x_1=0.5$. Solving $g_{x_1x_1x_1}=0$ explicitly in terms of $k$ is non-trivial. Instead, we examine its evolution as a function of $k$ in Fig.~S\ref{Figure_S3}.
The graphical analysis reveals the presence of an isolated zero for $k=k^*\simeq0.257$. We can now also easily verify that
$$ g_{x_1x_1x_1x_1x_1}(1/2,\mu^*,k^*,2.5,0.04)<0,\quad g_{x_1\mu}(1/2,\mu^*,k^*,2.5,0.04)>0,$$
which completes the recognition problem for the symmetric quintic bifurcation. 

\section{Oscillatory behavior from symmetric quintic bifurcation}

Unfolding the symmetric quintic bifurcation illustrated in Fig.~6 in the main text, and proved in Section~\ref{proof} in the direction of the subcritical pitchfork, introduces a new kind of multi-stability and with it a hysteresis between the distributed solution and the dominant solutions.  This allows for fast switching behavior in the $x_i$ for the closed-loop system with $N=2$ described by Eqs.~(1)-(4) from the main text even as $\mu$ increases and decreases slowly through the bifurcation point.  

In the case of $N=2$, $b_{11} = b_{22} = 1$, and $b_{12} = b_{21} = b \in (0,1)$, Eqs.~{(1)-(4) specialize to
\begin{eqnarray} 
\dot x_1&=& x_1(\sigma_{\gamma,k}(x_1)+b\sigma_{\gamma,k}(1-x_1))(1-\mu)+(1-x_1)(b\sigma_{\gamma,k}(x_1)+\sigma_{\gamma,k}(1-x_1))\mu-\phi x_1 \nonumber\\
\dot \mu &=& - K (x_1-\alpha_1)(\alpha_2-x_1) \mu (1-\mu). \label{EQ:reduced model}
\end{eqnarray}
When the two thresholds $x_1=\alpha_1$ and $x_1=\alpha_2$ intersect the critical manifold
$$\mathcal M_0=\{(x_1,\mu):\,x_1(\sigma_{\gamma,k}(x_1)+b\sigma_{\gamma,k}(1-x_1))(1-\mu)+(1-x_1)(b\sigma_{\gamma,k}(x_1)+\sigma_{\gamma,k}(1-x_1))\mu-\phi x_1=0\} $$
along its unstable branches, the closed-loop system (S\ref{EQ:reduced model}) exhibits two distinct oscillatory behaviors (Fig.~S\ref{Figure_S4}). Each of them corresponds to the periodic alternation between the distributed solution and one of the strongly dominant solutions. The associated limit cycles are symmetric with respect to the line $x_1=0.5$. They exist  for sufficiently small $K$.

The proof of existence and stability of these cycles uses techniques from geometric singular perturbations and blow-up theory \cite{druet2004blow}. In Fig.~S\ref{Figure_S5} we provide intuition on how these cycles can be geometrically constructed in the limit $K\to0$ by showing the singular ($K\to0$) phase portrait. Because $x_1$ is much faster than $\mu$, trajectories spend most of the time on the critical manifold. The singular limits of the two limit cycles are constructed as closed singular trajectories, which merge along the horizontal part of the critical manifold where $x_1=0.5$. Because $\mu$ is decreasing in that region, both singular cycles approach the vertical line $\mu=0$. There, they split in upward and downward directions, respectively. At the intersection with the upper and lower branches of the critical manifold, where $\mu$ is increasing, they slide along the critical manifold until the fold singularity, where they jump back to the horizontal branch of the critical manifold, much in the same way as a standard relaxation oscillator.

For small $K>0$ the two singular cycles perturb to two exponentially stable limit cycles, corresponding to the two oscillatory behaviors in Fig.~S\ref{Figure_S4}. By Fenichel theory \cite{fenichel1979geometric}, these two cycles are $\mathcal O(e^{-1/K})$-close to each other in the region where they shadow the horizontal branch of the critical manifold. It follows that, for $K$ sufficiently small, tiny perturbations, such as the noise described in the main text, make it possible for the system to switch between the two cycles. 

\begin{figure}[h!]
\centering
\includegraphics[width=0.50\textwidth]{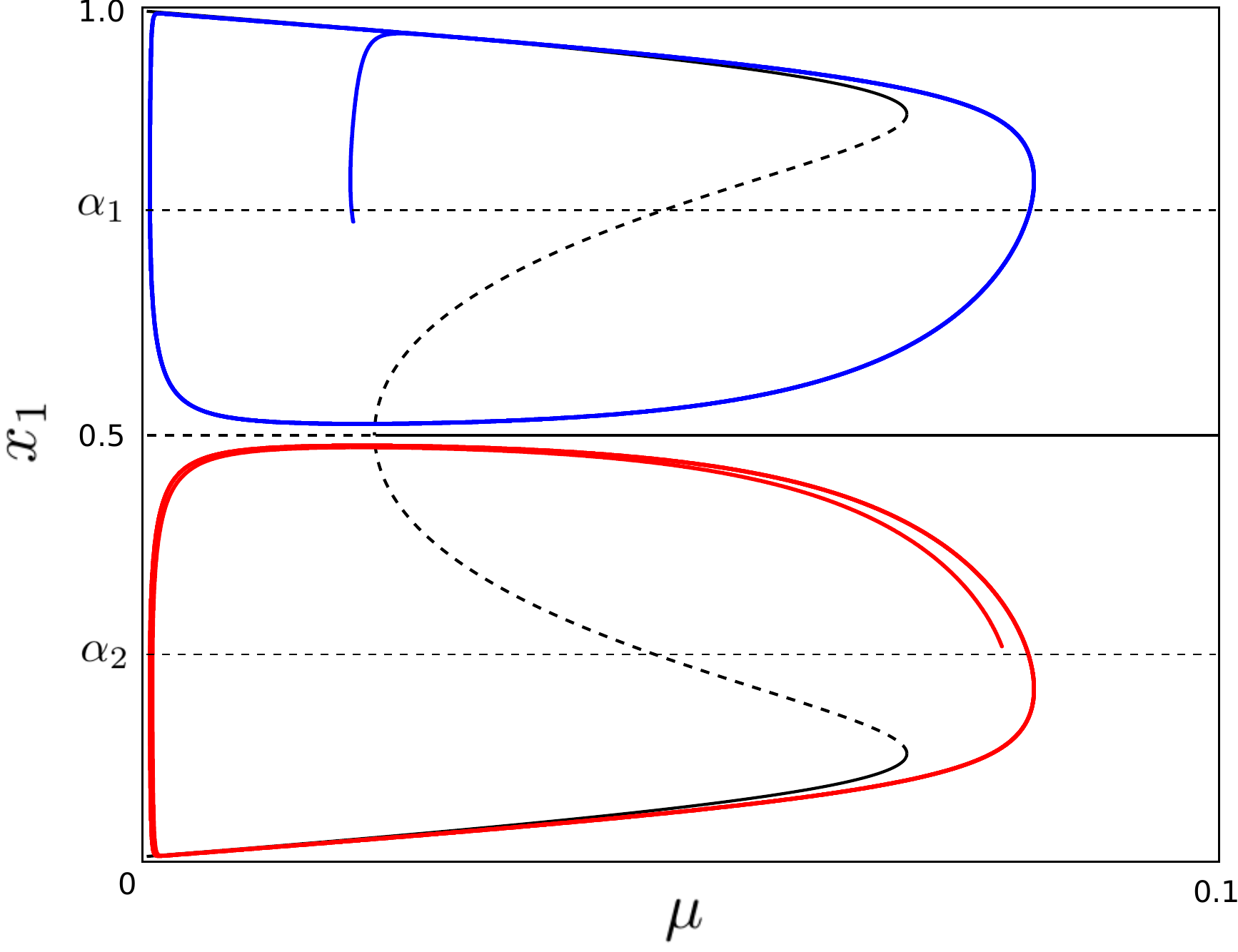}
\caption{Co-existence of two stable limit cycles (red and blue), shown on the bifurcation diagram for dynamics of Eq.~(S\ref{EQ:reduced model}). Stable manifolds are solid lines and unstable manifolds are dashed lines.  Parameters are $\gamma=2.5$, $k=0.01$, $K=0.3$, $\alpha_1=0.25$, and $\alpha_2=0.75$.}
\label{Figure_S4}
\end{figure}

\begin{figure}[htbp]
\centering
\includegraphics[width=0.50\textwidth]{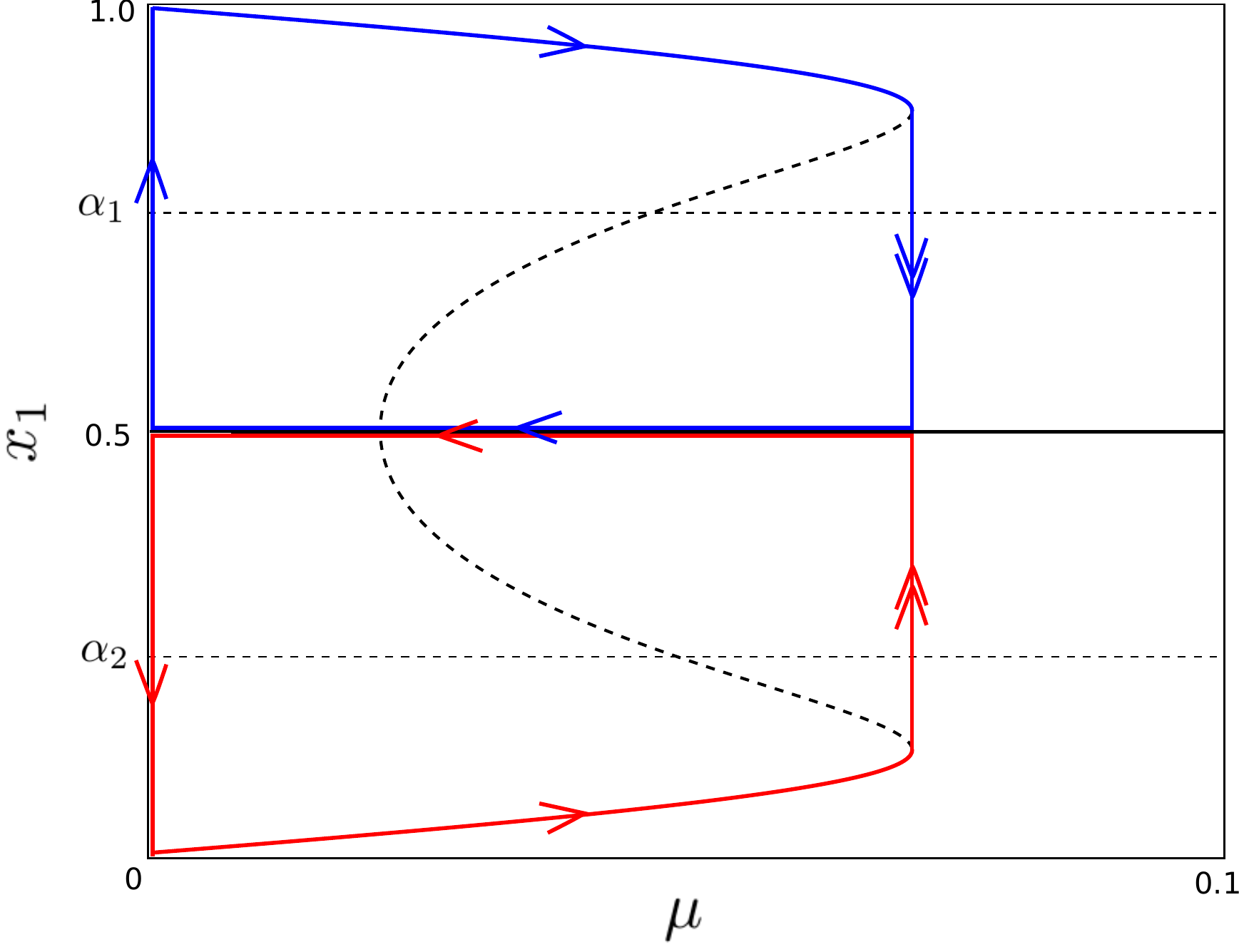}
\caption{Geometric construction of the two stable limit cycles (red and blue) in the limit $K\to0$, shown on the bifurcation diagram for dynamics of Eq.~(S\ref{EQ:reduced model}). Stable manifolds are solid lines and unstable manifolds are dashed lines.  Parameters are $\gamma=2.5$, $k=0.01$, $\alpha_1=0.25$, and $\alpha_2=0.75$.}
\label{Figure_S5}
\label{oscillations_quintic}
\end{figure}

\end{document}